%% file: main.tex
\documentclass[lettersize,journal]{IEEEtran}
\IEEEoverridecommandlockouts

\usepackage{cite}
\usepackage{amsmath,amssymb,amsfonts}
\usepackage{algorithmic}
\usepackage{graphicx}
\usepackage{amsfonts}
\usepackage{amssymb}
\usepackage{textcomp}
\usepackage{bm}
\usepackage{multirow} 
\usepackage{supertabular}
\usepackage[T1]{fontenc}
\usepackage{longtable} 
\usepackage{supertabular} 
\usepackage{xcolor}
\usepackage{subcaption}
\usepackage{tabularx}
\usepackage{comment}
\usepackage{booktabs}
\usepackage{mathrsfs} 
\usepackage{float} 
\usepackage{booktabs}

\DeclareUnicodeCharacter{00A0}{ }

\allowdisplaybreaks
\def\BibTeX{{\rm B\kern-.05em{\sc i\kern-.025em b}\kern-.08em
    T\kern-.1667em\lower.7ex\hbox{E}\kern-.125emX}}
\begin{document}

\title{Bi-Level Control of Weaving Sections in Mixed Traffic Environments with Connected and Automated Vehicles}

\author{Longhao Yan, Jinhao Liang, Kaidi Yang
\thanks{The authors are with the Department of Civil and Environment Engineering, National University of Singapore, Singapore 117576. Email:{longhao.yan@u.nus.edu, jh.liang@nus.edu.sg, kaidi.yang@nus.edu.sg}. \textit{Corresponding author}: Kaidi Yang. }
\thanks{This research was supported by the Singapore Ministry of Education (MOE) under NUS Start-Up Grant (A-8000404-01-00).
}}
\maketitle

\begin{abstract}
Connected and automated vehicles (CAVs) can be beneficial for improving the operation of highway bottlenecks such as weaving sections. This paper proposes a bi-level control approach based on an upper-level deep reinforcement learning controller and a lower-level model predictive controller to coordinate the lane-changings of a mixed fleet of CAVs and human-driven vehicles (HVs) in weaving sections. The upper level represents a roadside controller that collects vehicular information from the entire weaving section and determines the control weights used in the lower-level controller. The lower level is implemented within each CAV, which takes the control weights from the upper-level controller and generates the acceleration and steering angle for individual CAVs based on the local situation. 
The lower-level controller further incorporates an HV trajectory predictor, which is capable of handling the dynamic topology of vehicles in weaving scenarios with intensive mandatory lane changes. The case study inspired by a real weaving section in Basel, Switzerland, shows that our method consistently outperforms state-of-the-art benchmarks. 
\end{abstract}


\section{Introduction}
\IEEEPARstart{W}{eaving} sections are highway segments where merges are located close to diverges. Due to intense lane-changing maneuvers between the auxiliary and main lanes, weaving sections are recognized as key highway bottlenecks, characterized by a phenomenon called capacity drop. Such a capacity drop phenomenon has been widely reported in the literature, whereby the discharge flow of the weaving section can be reduced by 3\%-20\% after the onset of congestion~\cite{lertworawanich2003methodology,chen2018capacity}, further deteriorating highway traffic operations. 
Although researchers have identified potential causes of capacity drop (e.g., waste in space, undesirable distribution of lane-changing positions, etc.) and proposed various methods to mitigate capacity drop~\cite{tilg2018evaluating, yang2017feedback}
, these works focus on scenarios with only human-driven vehicles (HVs), which might be ill-informed due to limited perception capability~\cite{wortelen2013dynamic} and act independently with minimal coordination~\cite{kauffmann2018makes}.  

The development of connected and automated vehicles (CAVs) holds promise in improving traffic operations in weaving sections~\cite{cai2022formation, han2022strategic}. 
Enabled by vehicle-to-vehicle (V2V) and vehicle-to-infrastructure (V2I) wireless communications, CAVs can provide holistic and predictive information about the weaving section. Moreover, unlike HVs that are controlled by independent drivers, CAVs can accurately follow instructions provided by a central controller and thus can be coordinated to improve system-wide efficiency in weaving sections. 

To exploit the benefits of CAVs in highway systems, researchers have proposed various methods to control lane-changing maneuvers~\cite{wang2022ego,wu2023integrated}. 
Most of these works assume no coordination or only local coordination that only coordinates a vehicle with its neighbors~\cite{peng2022integrated,zhang2022multi}. 
For example, Ref. \cite{wang2021harmonious} proposed a harmonious deep reinforcement learning (DRL)-based algorithm to facilitate lane changing for CAVs with limited perception range, whereby all vehicles act simultaneously to make independent decisions to balance system and individual efficiency. Ref. \cite{peng2022integrated} developed DRL algorithms to make car-following and lane-changing decisions for a single vehicle while no coordination is applied among vehicles. Ref. \cite{zhang2022multi} proposed a multi-agent DRL-based lane-changing control approach in which vehicles are controlled in a decentralized manner using only information collected by themselves. 
Although these local coordination approaches can be efficient in improving highway operations by involving only a few vehicles, they only look at a small portion of highway segments and hence ignore the global impact of the resulting local decisions on the entire weaving section, making it hard to achieve system-wide efficiency. In fact, global coordinated control is argued to be essential for improving traffic efficiency in weaving sections where mandatory lane changes are prevalent~\cite{tajalli2022distributed,yang2022coordinated}.

To the best of our knowledge, only a small portion of works have developed \emph{global coordination} methods that jointly control vehicles across weaving sections~\cite{han2022strategic, amini2021optimizing}. 
Ref. \cite{yang2022coordinated} proposed a dynamic programming algorithm to coordinate the trajectory planning of all CAVs in scenarios with intensive mandatory lane-changing maneuvers. Ref. \cite{zhang2022hybrid} developed a hybrid Model Predictive Control (MPC) method for systematic cooperative lane-changing control to improve traffic smoothness and driving comfort. A distributed cooperative lane-changing method was developed by~\cite{tajalli2022distributed}, which utilized the predictive information from other vehicles to solve a mixed integer optimization problem. However, these aforementioned works suffer from two major limitations. First, they tend to involve solving complex optimization problems, which do not scale well as the number of vehicles increases.
Second, most of these works assume all vehicles to be CAVs without considering the transition period in which CAVs are gradually developed and deployed. Although Ref. \cite{tilg2018evaluating} considered such a transition period, this work leveraged a macroscopic traffic modeling approach to derive the optimal distribution of lane-changing positions without coordinating individual CAVs.  

To address the issues with existing local and global coordination methods, we propose a computationally efficient bi-level controller that combines DRL and MPC to globally coordinate lane-changing decisions of all CAVs in mixed-traffic weaving scenarios. \vspace{0.15em}

\noindent \emph{Statment of Contribution}.
Our contribution is two-fold. First, we propose a bi-level controller with an upper-level DRL and a lower-level MPC for the global coordination of CAVs at weaving sections in mixed-traffic environments. The upper-level DRL-based controller represents a roadside coordinator that collects information from each vehicle to decide the weights of multiple control objectives to inform the lower-level controller. The lower-level MPC-based controller is implemented in each vehicle, which takes decisions from the upper-level controller and generates driving strategies for individual CAVs according to the local situation. Our bi-level approach can enhance system-wide efficiency while allowing individual vehicles to react to changing local situations considering safety and adaptability in mixed-traffic scenarios. Second, unlike a large number of MPC-based controllers that assume constant speed or pre-known trajectory for surrounding HVs~\cite{qiu2023cooperative,sun2024multi}, at the lower level, we design an HV trajectory prediction algorithm that can predict the trajectories of multiple HVs around a CAV, providing necessary input to the lower-level MPC-based controller. Specifically, we leverage EvolveGCN~\cite{pareja2020evolvegcn} to consider the evolution of the dynamic topology of surrounding HVs around CAV.  

The remainder of this paper is organized as follows. Section~\ref{sec:Problem Formulation and Overall Framework} describes the problem formulation and overall framework of the proposed bi-level controller. Section~\ref{sec:Upper_Level_Controller} describes the upper-level DRL-based controller, and Section~\ref{sec:Lower_Level_Controller} elaborates the formulation of MPC-based controller and the details of HV trajectory predictor. Section~\ref{sec:Result} performs the simulation of a typical weaving section. Section~\ref{sec:Conclusion} concludes the paper and presents future work.

\section{Problem Formulation and Overall Framework} \label{sec:Problem Formulation and Overall Framework}
\begin{figure}[htbp]
	\centering
	\includegraphics[width=0.48\textwidth]{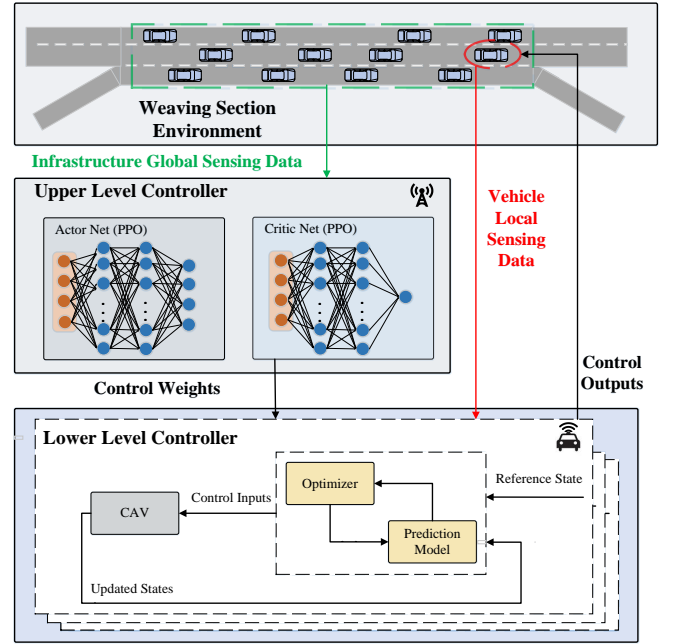}
	\caption{Framework of the global coordinated bi-level controller in weaving sections.}
	\label{fig:overallframework}
\end{figure}

We consider a typical weaving section of $L$ lanes with mixed traffic of CAVs and HVs, whereby CAVs are controlled by their onboard controllers, and HVs are controlled by drivers. Both CAVs and HVs are assumed to be connected and can communicate with a roadside unit installed near the weaving section through V2I communication. We make this assumption because we envision that vehicle connectivity will be prevalent when automated vehicles are deployed. In fact, the Land Transport Authority in Singapore is mandating Singapore-registered vehicles to install onboard units, making almost all vehicles connected in the near future. Mathematically, let $\mathcal{C}$ represent the set of CAVs, and $\mathcal{H}$ represent the set of HVs. The state of each vehicle $i \in \mathcal{C}\cup\mathcal{H}$ (either CAV or HV) is represented by a tuple $s_i = (e_i, \boldsymbol{p}_i, v_i, \psi_i)$, with $e_i$ indicating its destination lane, $\boldsymbol{p}_i = (p_{x,i}, p_{y,i})$ the position vector containing the longitudinal coordinate $p_{x,i}$ and the lateral coordinate $p_{y,i}$, $v_i$ the longitudinal speed, and  $\psi_i$ the heading angle. 
The considered time horizon is discretized into a set $\mathcal{K}$ of time intervals (indexed by $k$) with size $\Delta t$. 

The goal of this work is to develop a global coordination algorithm for CAVs to improve the operation of the weaving sections in mixed traffic. 
To this end, we propose a bi-level control approach, as demonstrated in Fig.~\ref{fig:overallframework}, whereby the roadside unit implements an RL-based upper-level controller (see Section~\ref{sec:Upper_Level_Controller}), and each CAV implements an MPC-based lower-level controller (see Section~\ref{sec:Lower_Level_Controller}). 
The upper-level controller collects information from all vehicles on the road and then determines the weights of various control objectives (hereafter referred to as control weights) in lower-level controllers to optimize system performance. 
The lower-level controller embedded in each CAV first predicts the trajectories of the surrounding HVs, if any, and then determines the ego CAV's acceleration rate and steering angle to optimize various control objectives, such as reaching the target, traffic efficiency, and safety, using the control weights calculated at the upper level. 
The proposed bi-level controller can improve mixed-traffic weaving section operations by achieving the global coordination of CAVs in a computationally efficient manner while providing local safety guarantees to CAVs. Note that for presentation simplicity, we assume both levels make decisions at the same frequency (i.e., with the same time interval $\Delta t$). This assumption can be readily relaxed such that the upper-level controller makes decisions at a lower frequency than the lower-level controller (see the sensitivity analysis in Section~\ref{sec:sens_cell}). 

We further make two assumptions regarding the HV driving behavior, characterized by car-following and lane-changing models. 
First, we make a realistic assumption that the HV driving behavior is heterogeneous and unknown to CAVs. Second, we assume that HVs do not adapt their driving strategies to the real-time control of CAVs because such adaptation can only happen slowly after HVs obtain sufficient observations about CAV behavior. 

\section{RL-based Upper-level Controller for Global Coordination} \label{sec:Upper_Level_Controller}
We formulate the decision-making process of the upper-level controller (i.e., the roadside controller) as a Markov Decision Process (MDP) $\mathcal{T} = (\mathcal{S}, \mathcal{A}, \mathcal{P}, R, \gamma)$, where $\mathcal{S}$ represents the state space, 
$\mathcal{A}$ the action space, $\mathcal{P}$ system dynamics, $R$ reward function, and $\gamma$ a scalar discount factor. \vspace{0.3em}

\noindent \emph{State Space $\mathcal{S}$}. \label{sec:state}
Let us divide the lanes of the weaving section into a set of cells, collectively denoted by $\mathcal{N}$ and indexed by $n$. 
We define the states at time step $k\in\mathcal{K}$ as 
$\bm{s}_k=\{g_{n,k}\}_{n\in\mathcal{N}}$, where $g_{n,k}$ represents the state of cell $n\in\mathcal{N}$ of the past $D$ time steps, represented as $g_{n,k} = \{l_{n, k-h}, b_{n, k-h}, \bar{v}_{n, k-h}\}_{h=0}^{D-1}$, which includes the lane change occupancy $l_{n, k-h}$ (i.e., the ratio of vehicles not on the destination lane), the number of vehicles located in the cell $b_{n, k-h}$, and the average speed $\bar{v}_{n, k-h}$ of all vehicles within cell $n$ at time step $k-h$. 
\vspace{0.3em}

\noindent \emph{Action Space}. \label{sec:action} 
We define the upper-level action as 
$\boldsymbol{\alpha} = \{\alpha_{n,k}\}_{n\in\mathcal{N}}$, where $\alpha_{n,k} = \{\alpha_{n,k}^{\text{effort}}, \alpha^{\text {exit,x}}_{n,k}, \alpha^{\text {exit,y}}_{n,k}, \alpha^{\text {vel}}_{n,k}, \alpha^{\text{acc}}_{n,k}\}$ represents the control weights in the MPC-based lower-level controller (see Section~\ref{sec:Lower_Level_Controller} for details) for all CAVs in cell $n$, which penalizes the control effort ($\alpha_{n,k}^{\text{effort}}$), the distance to the target point ($\alpha^{\text {exit,x}}_{n,k}$ and $\alpha^{\text {exit,y}}_{n,k}$), the deviation from the desired speed ($\alpha^{\text {vel}}_{n,k}$), and the impact on surrounding vehicles ($\alpha^{\text{acc}}_{n}$). 
We make two remarks about using these control weights as upper-level actions. First, these weights are used by every CAV in cell $n$ to trade off these control objectives to adapt the lower-level controller to various traffic conditions. Second, these control weights can help achieve global coordination in that they can encode the global information about the weaving section through the policy network trained in the RL-based upper-level controller. \vspace{0.3em}

\noindent \emph{System Dynamics}. We describe the system dynamics $\mathcal{P}$ as the traffic dynamics of the weaving section, including the arrival and departure of vehicles, the car-following and lane-changing decisions of the HVs, and the kinematic models of both the HVs and CAVs. Note that the system dynamics are unknown from the perspective of the upper-level RL controller. \vspace{0.3em}

\noindent \emph{Reward Function}. \label{sec:reward} 
We define the reward $r_{k}^{\text{total}}$ at time step $k$ as the weighted sum of three components, including (i) a speed component $r_{k}^{\text{spd}}$, (ii) a traffic flow component $r_{k}^{\text{flow}}$, and (iii) a missing penalty component $r_{k}^{\text{miss}}$ penalizing vehicles that fail to perform lane changes at the end of the weaving section:  
\begin{align}
r_{k}^{\text{total}} &= \beta^{\text{spd}}r_{k}^{\text{spd}} + \beta^{\text{flow}}r_{k}^{\text{flow}} + \beta^{\text{miss}}r_{k}^{\text{miss}} 
\end{align}
where $\beta^{\text{spd}}$, $\beta^{\text{flow}}$, and $\beta^{\text{miss}}$ are the corresponding weights defined by the traffic operator. The specific forms of these three reward components are given below: 
\begin{align}
r_{k}^{\text{spd}} = \frac{1}{N_k} \sum_{i = 1}^{N_k} \frac{v_{i,k}}{v_{\text{f}}}, \quad r_{k}^{\text{flow}} = \frac{q_{D,k}}{q_{\text{max}}} , \quad 
r_{k}^{\text{miss}} = b_{k}^{\text{miss}} 
\end{align}
where the speed component $r_{k}^{\text{spd}}$ represents the normalized space-mean speed on the weaving section, with $N_k$, $v_{i,k}$, and  $v_{\text{f}}$ indicating the number of vehicles on the weaving section at time step $k$, the speed of vehicle $i$  at time step $k$, and the free-flow speed, respectively. Notice that the calculation of $r_{k}^{\text{spd}}$ considers both CAVs and HVs to ensure that the decisions made by RL benefit the entire traffic flow regardless of vehicle type.  
The traffic flow component $r_{k}^{\text{flow}}$ represents the exit flow $q_{D,k}=\{q_{k-h}\}_{h=0}^{D-1}$ of the weaving section in the past $D$ time steps, normalized by the theoretical capacity $q_{\text{max}}$.  Here, the theoretical capacity refers to the capacity without considering lane changing, as frequent lane changes in weaving sections can lead to capacity drop. 
The missing penalty component
$r_{k}^{\text{miss}}$ is defined to be the number of vehicles that leave the weaving section without successfully changing to their destination lane at time step $k$, represented by $b_{k}^{\text{miss}}$.\vspace{0.3em}

\noindent \emph{Deep Reinforcement Learning Algorithm}. \label{sec:drl} 
The formulated MDP is solved using a DRL algorithm, Proximal Policy Optimization (PPO), which is selected due to its stability and wide application in traffic and vehicle control ~\cite{wang2022ego,gu2022integrated}. 
PPO leverages an actor-critic structure, whereby the actor network takes the grid-like global information of the weaving section as input and provides an output of the control weights for the lower-level MPC controller, and the critic network takes the same input but outputs the estimated state value of MDP. Both the actor and the critic network use Convolutional Neural Networks (CNN) as input layers to handle grid-like states~\cite{li2023coor}. 

The actor network determines a policy $\pi_{\theta}(\bm{\alpha}\mid\bm{s})$ parameterized by $\theta$ to minimize a clipped loss represented in Eq.(\ref{eq:actor_loss}), where a clip function with threshold parameter $\kappa$ is used to avoid excessive updates that could result in performance oscillations: 
\begin{align}
\begin{split}
L_{\mathrm{CLIP}}(\theta) & = {\hat{\mathbb{E}}}\left[\min \left( \hat{{A}_{t}}\frac{\pi_{\theta}(\bm{\alpha}\mid\bm{s})}{\pi_{\theta_{\text {old }}}(\bm{\alpha}\mid\bm{s})}, \right. \right. \\
& \left. \left. \hat{{A}_{t}}\operatorname{clip}\left(\frac{\pi_{\theta}(\bm{\alpha}\mid\bm{s})}{\pi_{\theta_{\text {old }}}(\bm{\alpha}\mid\bm{s})}, 1-\kappa, 1+\kappa\right) \right)\right]
\end{split}
\label{eq:actor_loss}
\end{align}
where $\pi_{\theta_{\text {old }}}$ is an old policy updated regularly to confine the variance of the current policy, and $\hat{A_{t}}$ is an advantage function to accelerate convergence. 

The critic network estimates the state value $V_{\mu}(s_{t})$ of the MDP. The loss function of critic net with parameter $\mu$ is defined as:
\begin{align}
L^{V}(\mu) & = \hat{{\mathbb{E}}}\left[\left|\hat{V}_{\mu}^{\text {target }}\left(s_{t}\right)-V_{\mu}\left(s_{t}\right)\right|\right] \label{eq:critic}
\end{align}
where $\hat{V}_{\mu}^{\text {target }}\left(s_{t}\right)=r_{t+1}+\gamma V_{\mu}\left(s_{t+1}\right)$ is the state value that $V_{\mu}(s_{t})$ aims to approach. 

The actions determined by the DRL-based upper-level controller will be leveraged to inform the MPC-based lower-level controller, as presented next. 

\section{MPC-based Lower-Level Controller for Individual Vehicle} \label{sec:Lower_Level_Controller}
In this section, we first introduce our proposed lower-level MPC controller that combines behavioral decision-making, path planning, and motion control of individual CAVs in Section~\ref{sec:MPC_formulation}. Then, we present our HV trajectory prediction algorithm to generate input for the MPC in Section~\ref{sec:HV_predictor}.

\subsection{MPC Formulation} \label{sec:MPC_formulation}
In this subsection, we present the lower-level controller for CAV $i\in\mathcal{C}$ with its center of gravity (COG) located within cell $n_i$ on lane $o_{i,k}$ with destination lane $e_i$. 
For brevity, we omit the index $i$ in this subsection where its omission does not introduce ambiguity. 
At each time step $k$, this CAV $i$, referred to as the subject vehicle (SV), determines its optimal lane-changing decision, acceleration rate, and steering angle using two types of input: (i) the detected states of its surrounding vehicles within a sensing range and (ii) the control weights associated with the cell in which it is located, which is received from the upper-level controller. These control decisions are made through an MPC framework, where optimization is performed to minimize total costs within a planning horizon of the next $P$ time steps, but only the one corresponding to the current time step is executed. The embedded optimization problem of the MPC framework is a convex optimization problem that minimizes a combination of control objectives: 
\begin{align}
    J &=  \alpha_{n,k}^{\text{effort}} J^{\text{effort}}_k + J^{\text{safety}}_{k} + \alpha_{n,k}^{\text{exit,x}} J^{\text{exit,x}}_{k}  \notag \\ 
     &\quad + \alpha_{n,k}^{\text{exit,y}} J^{\text{exit,y}}_k+ \alpha_{n,k}^{\text{vel}} J^{\text{vel}}_k + \alpha_{n,k}^{\text{acc}}J^{\text{acc}}_k \label{eq:obj}
\end{align}
where $J^{\text{effort}}_k$, $J^{\text{safety}}_k$, $J^{\text{exit,x}}_k$, $J^{\text{exit,y}}_k$, $J^{\text{vel}}_k$, and $J^{\text{acc}}_k$, respectively, represent the control effort, safety penalty, longitudinal distance to the target point, lateral distance to the target point, deviation from the desired speed, and the impact on the surrounding vehicles at time step $k$, weighted by the control weights (relative to the safety penalty) obtained from the upper-level controller $\alpha_{n,k} = \{\alpha_{n,k}^{\text{effort}}, 1, \alpha^{\text {exit,x}}_{n,k}, \alpha^{\text {exit,y}}_{n,k}, \alpha^{\text {vel}}_{n,k}, \alpha^{\text{acc}}_{n,k}\}$.\vspace{0.3em} 

\noindent \emph{Notation}. We use $z(k)$ to represent the value of variable $z$ at time step $k$ observed in the real system, $z(k+t|k)$ to represent the predicted value of variable $z$ at time step $k+t$ with the prediction being made at time step $k$, and $\bm{z}_k=\{z(k+t|k)\}_t$ to represent the vector of variables formed by $z(k+t|k)$.\vspace{0.3em}  

Next, we present the details of these control objectives and the corresponding constraints. 

\vspace{0.3em}
\noindent \emph{(1) System Dynamic} \label{sec:sys_dyn}
\vspace{0.3em} 

We first describe the vehicle dynamics and obtain the control objective of the control effort $J^{\text{effort}}_k$. For vehicle dynamics, we employ the widely adopted bicycle model~\cite{gong2023game} to describe the dynamics of longitudinal position $p_{x}$, lateral position $p_{y}$, heading angle $\psi$, and longitudinal speed $v$ of SV.
\begin{subequations}
\begin{align}
\dot{p_{x}} & = v  \label{eq:dyn_x} \\
\dot{p_{y}} & = v \psi+\frac{l_{{r}}}{l_{f}+l_{r}} v \delta_{f} \\
\dot{\psi} & = \frac{l}{l_{{f}}+l_{r}} v \delta_{f} \\
\dot{v} & = a_{x} \label{eq:dyn_acc}
\end{align}
\end{subequations}
where $a_x$ and $\delta_{f}$ are the longitudinal acceleration and steering angle, respectively, serving as the decision variables. 
$l, l_{f}$, and $l_{r}$ represent the length of the axle, the distance of the front axle from the COG, and the distance of the rear axle from the COG, respectively.

To facilitate the design of MPC, we linearize and discretize the bicycle model Eq.(\ref{eq:dyn_x})-Eq.(\ref{eq:dyn_acc}) into a discrete-time linear time-invariant system as follows.
\begin{align}
\bm{x}(k+t+1|k) & = \bm{A} \bm{x}(k+t|k)+\bm{B} \bm{u}(k+t|k) \label{eq:system_dynamics}
\end{align}
where the state variable $\bm{x}(k+t|k) = [p_{x}(k+t|k), p_{y}(k+t|k), \psi(k+t|k), v(k+t|k)]$, control variable $\bm{u}(k+t|k) = [a_{x}(k+t|k), \delta_{f}(k+t|k)]$. Matrices $\bm{A}$ and $\bm{B}$ describe the linearized system dynamics. 

Furthermore, we link SV's lateral positions with the lane-changing decisions. Since intensive lane changes can reduce traffic efficiency, we only allow SV to perform mandatory lane changes towards its destination lane $e$. Moreover, since consecutive lane changes within a short time period are typically not allowed, we assume that during the planning horizon, SV can only change to the adjacent lane in the direction of its destination lane $e$. 
In particular, we use a binary variable $\varsigma_{k}\in \{0,1\} $ to indicate whether lane changes will happen during the planning horizon starting at time step $k$, with $\varsigma_{k}=1$ indicating lane changing. Accordingly, the above requirements can be mathematically represented as follows: 
\begin{align}
    &\varsigma_{k} \leq |o_k-e| \label{eq:lc_not_on_des}\\
    & \varsigma_{k} \geq \varsigma_{k-1}(1-|o_k-o_{k-1}|)\label{eq:lc_complete}\\
     &p_y(k+P \mid k) = \left\{
    \begin{aligned}
        &p_{y,o_k}, &\text{if }o_k=e \\
        &\varsigma_{k}p_{y,o_k+1} + (1-\varsigma_{k})p_{y,o_k},&\text{if }o_k<e \\
        & \varsigma_{k}p_{y,o_k-1} + (1-\varsigma_{k})p_{y,o_k},&\text{if }o_k>e
    \end{aligned} 
    \right.  \label{eq:final_lateral_position} \\
    &\left\{
    \begin{aligned} 
        &p_{y,o_k}^-  \leq p_y(k+t \mid k) \leq p_{y,o_k}^+, &\text{if }o_k=e \\
        &p_{y,o_k}^-  \leq p_y(k+t \mid k) \leq \varsigma_{k}p_{y,o_k+1}^+ + (1-\varsigma_{k})p_{y,o_k}^+ ,&\text{if }o_k<e \\
        & \varsigma_{k}p_{y,o_k-1}^- + (1-\varsigma_{k})p_{y,o_k}^- \leq p_y(k+t \mid k)  \leq p_{y,o_k}^+,&\text{if }o_k>e
    \end{aligned} 
    \right.  \label{eq:range_lateral_position}
\end{align}
where Eq.(\ref{eq:lc_not_on_des}) represents that lane changing decision can only be made when the current lane $o_k$ does not coincide with the destination lane $e$ (i.e., $o_k\neq e$). Eq.(\ref{eq:lc_complete}) represents that if lane changing is made at the previous time step $k-1$ and the lane change is not completed, the lane changing process will continue at time step $k$. Eq.(\ref{eq:final_lateral_position}) calculates the lateral position at the end of the planned horizon,  $p_y(k+P \mid k)$, 
where $p_{y,o_k}$ represents the lateral position of the center line of lane $o_k$ on which SV is currently located. If SV decides not to change lane ($\varsigma_{k}=0$), its lateral position at the end of the prediction horizon, $p_y(k+P \mid k)$, would be the center of its current lane. If SV decides to change lane ($\varsigma_{k}=1$), its lateral position at the end of the prediction horizon, $p_y(k+P \mid k)$, would be the center of its adjacent lane ($p_{y,o_k+1}$ or $p_{y,o_k-1}$). Eq.(\ref{eq:range_lateral_position}) represents the range of the lateral positions of SV during the planning horizon (with $t=1,\cdots,P$), where $p_{y,\xi}^-$ and $p_{y,\xi}^+$ represent the lateral position of lane $\xi$'s borders. Specifically, the SV's lateral position during the planning horizon should be within the borders of its current lane 
if the lane-keeping decision is made ($\varsigma_k=0$) and between its current lane and its adjacent lane in the direction of its destination lane if the lane-changing decision is made ($\varsigma_k=1$).  

Next, we define the physical limitation constraints for other state and control variables of SV within the planned horizon. First, SV's speed should be upper-bounded by the speed limit on the weaving section. Second, SV's acceleration rate and steering angle are constrained by the operational limits defined by the vehicle's mechanics. Overall, these physical limitation constraints are summarized as follows:
\begin{align}
\bm{x}_{\text {min }} &\leq \bm{x}(k+t|k) \leq \bm{x}_{\text {max }} \label{eq:x_bound}\\
\bm{u}_{\text {min }} &\leq \bm{u}(k+t|k) \leq \bm{u}_{\text {max }} \label{eq:u_bound}
\end{align}
where $\bm{x}_{\text {min }}$ and $\bm{x}_{\text {max }}$ represent the lower and upper bounds for state variables, respectively, and $\bm{u}_{\text {min }}$ and $\bm{u}_{\text {max }}$ represent the lower and upper bounds for the control variables, respectively. 

Then, the control effort $J^{\text{effort}}_k$ describes the energy consumption associated with the control variables, represented as follows: 
\begin{align}
J^{\text{effort}}_k &= \sum_{t = 1}^{P} \|\bm{u}(k+t \mid k)\|_2^2
\end{align}

\vspace{0.3em}
\noindent \emph{(2) Driving Safety} \label{sec:safe_con}
\vspace{0.3em} 

\begin{figure}[h]
	\centering
	\includegraphics[width=0.4\textwidth]{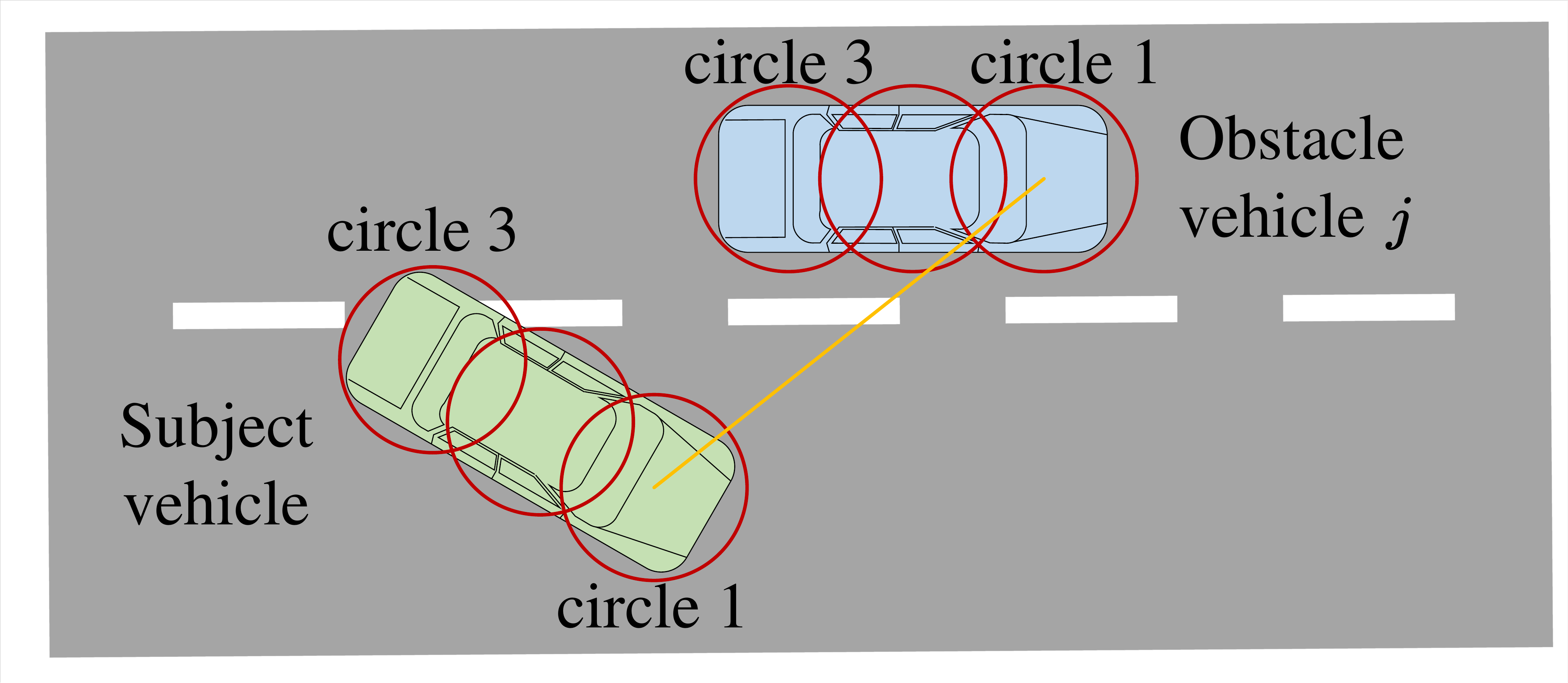}
	\caption{Illustration of circle decomposition.}
	\label{fig:cir_demo}
\end{figure}

We now establish the driving safety constraints and derive the safety penalty $J^{\text{safety}}_k$. Driving safety can be achieved by minimizing the reciprocal force between vehicles~\cite{liang2022mas}, 
defined as the safety violation over a safe distance between SV and its surrounding vehicles. To enhance vehicle spacing safety, we employ a three-circle decomposition method introduced in ~\cite{wang2021path} to refine the reciprocal force, as illustrated in Fig.~\ref{fig:cir_demo}.

As shown in Fig.~\ref{fig:cir_demo}, vehicles are decomposed into three circles whose centers are located at the center of the front axle, gravity, and back axle, respectively. This decomposition enhances the precision of the reciprocal force compared to relying solely on the COG. The objective of the reciprocal force is to maintain a safe distance between any pair of vehicle circles. Consequently, the corresponding constraint is expressed as follows:
\begin{align}
&\| \bm{p}^{\text{SV}}_{l}(k+t \mid k) -  \bm{p}^{\text{safety}}_{j, w}(k+t \mid k) \|_{2} \geq d^{\text{safety}}-\iota_{ljw}(k+t \mid k), \notag\\
&~~\forall j \in \mathcal{M}, l \in\{1,2,3\}, w \in\{1,2,3\}, t \in\{1, \ldots, P\}
\label{eq:const_rep}
\end{align}
where $\boldsymbol{p}_{l}^{\text{SV}}$ represents the center position of SV's $l$-th circle, and $\bm{p}^{\text{safety}}_{j, w}$ is the center position of $w$-th circle of surrounding vehicle $j \in \mathcal{M}$. 
Here, $\mathcal{M}$ represents the surrounding vehicles of SV, which comprise a maximum of four vehicles, including the leading and following vehicles in the current lane and those in the adjacent lane in the direction of SV's destination lane. 
Moreover, we only consider surrounding vehicles that SV can continuously observe for more than three seconds, this is a practical requirement because vehicles outside of three seconds are impossible to collide with SV. 
If the surrounding vehicle $j$ is an HV, the values of $\bm{p}^{\text{safety}}_{j, w}$ can be calculated from the positions of the surrounding vehicles over the planning horizon, which is predicted using the trajectory prediction algorithm presented in Section~\ref{sec:HV_predictor}. 
If surrounding vehicle $j$ is a CAV, the value of $\bm{p}^{\text{safety}}_{j, w}$ can be inferred from the control variables of vehicle $j$ at time step $k-1$ obtained through V2V communication. 
Constant $d^{\text{safety}}$ is a predefined safe distance threshold to avoid collisions, and $\iota_{ljw}(k+t \mid k)$ indicates the violation between SV's $l$-th circle and $w$-th circle of surrounding vehicle $j$. 

To make the resulting MPC formulation convex, we linearize Eq.~(\ref{eq:const_rep}) using its first-order Taylor expansion as suggested in~\cite{soria2021distributed}. 
\begin{align}
\boldsymbol{A}^{\text{safety}}_{k}\left[\left( \bm{U}_{k}\right)^{T}, \left( \bm{X}_{k}\right)^{T},\left(\bm{Z}_{k}\right)^{T}\right]^{T} &\leq \boldsymbol{b}^{\text{safety}}_{k} \label{eq:safety1}\\
-\bm{Z}_{k} &\leq 0 \label{eq:safety2}
\end{align}
where $\bm{U}_{k}=\{\bm{u}(k+t\mid k)\}_{t=0}^{P-1}$, $\bm{X}_{k}=\{\bm{x}(k+t\mid k)\}_{t=1}^{P}$, and $\bm{Z}_{k}=\{\iota_{ljw}(k+t \mid k)\}_{t\in\{1,\cdots,P\}, w,l \in \{1,2,3\}, j \in \mathcal{M}}$ are the vector of the control, state, and slack variables, respectively. Matrix $\boldsymbol{A}^{\text{safety}}$ and vector $\boldsymbol{b}^{\text{safety}}$ define the linearized constraint.

Then, the reciprocal force is formulated as a quadratic function representing the violation of the safe distance, which is defined as follows:
\begin{align}
J^{\text{safety}}_k &= \sum_{t = 1}^{P} \chi^{t}\|\zeta_{t}\|_2^2
\end{align}
where $\zeta_{t} = \sum_{l=1}^{3}\sum_{w=1}^{3}\sum_{j \in \mathcal{M}}\iota_{ljw}(k+t \mid k)$ represents the cumulative violations of safe distance between SV and its surrounding vehicles in the set $\mathcal{M}$. The decay factor $\chi$ is designed to highlight the risks of immediate collisions. 

\vspace{0.3em}
\noindent \emph{(3) Exit Attraction} \label{sec:destination_con}
\vspace{0.3em} 

Next, we calculate $J^{\text{exit,x}}_k$ and $J^{\text{exit,y}}_k$, which represent the exit attraction defined as the vehicle's desire to approach its destination exit point. Such criteria are especially important for weaving sections where most lane changes are mandatory. As the SV approaches the end of the weaving section, its urgency in performing lane change increases. Consequently, we can define the cost terms $J^{\text{exit,x}}_k$ and $J^{\text{exit,y}}_k$ as follows: 
\begin{subequations}
\begin{align}
J^{\text{exit,x}}_k = \left(p_x(k+P \mid k)-\hat{p}_{x,k}\right)^2 \\
J^{\text{exit,y}}_k = \left(p_y(k+P \mid k)-p_{y,\text{d}}\right)^2
\end{align}
\end{subequations}
where $p_{y,d}$ represents the coordinate of the lateral center of the destination lane, and $\hat{p}_{x,k}$ represents SV's estimated longitudinal position at the end of the prediction horizon. The value of $\hat{p}_{x,k}$ is estimated by assuming a free-flow speed $v_{\text{f}}$ throughout the prediction period. 

\vspace{0.3em}
\noindent \emph{(4) Target Speed Attraction} \label{sec:APF_vel}
\vspace{0.3em} 

To encourage the SV to maintain a high speed, an attractive force toward the target speed is defined as follows:
\begin{align}
J^{\text{vel}}_k = (v_{k}-v_{\text{f}})^{2} \label{eq:cost_vel}
\end{align}
where the target speed is set to the free-flow speed $v_{\text{f}}$ along the weaving section.

\vspace{0.3em}
\noindent \emph{(5) Impact on SV and Other Vehicles} \label{sec:impact_vehicles}
\vspace{0.3em} 

We finally account for the impact of SV's lane change on surrounding vehicles to obtain $J^{\text{acc}}_k$ using an incentive term inspired by the Minimize Overall Braking Induced by Lane Change (MOBIL) model~\cite{Kesting2007GeneralLM}, which calculates the impact of SV's lane changing decision on itself and surrounding vehicles: 
\begin{align}
B_{k}^{\text{acc}} = \underbrace{a_{i,k}-\hat{a}_{i,k}}_{\text{SV}}+\rho (\underbrace{a_{f,k}-\hat{a}_{f,k}}_{\text{new follower }}+\underbrace{a_{f\prime,k}-\hat{a}_{f\prime,k}}_{\text{old follower}}) \label{eq:Atr_acc}
\end{align}
where $a$ and $\hat{a}$ represent longitudinal acceleration rates before and after lane change, respectively, and the subscripts $i,~f$, and $f\prime$ indicate SV, the new follower after lane change, and the old follower before lane change, respectively. $\rho$ is the factor called politeness that represents the level of consideration of the surrounding vehicles.

To estimate the acceleration after lane change, we leverage the Intelligent Driver Model (IDM)~\cite{albeaik2022limitations}, which models the longitudinal behavior of a vehicle as:
\begin{align}
\dot{v} & =a_{\text{max}}\left[1-\left(\frac{v}{v_{\text{f}}}\right)^{4 }-\left(\frac{d^{*}}{d}\right)^{2}\right] \label{eq:IDM1} \\
d^{*} & =d_{\text{jam}}+\tau v+\frac{v \Delta v}{2 \sqrt{a_{\text{max}} a_{\text{min}} }} \label{eq:IDM2}
\end{align}
where $v$ is the vehicle speed, $d$ is the spacing, and $\Delta v$ is the speed difference between the vehicle and its leading vehicle. Parameter $\tau$ is the desired time headway, $d_{\text{jam}}$ is the jam spacing,  and $a_{\text{max}}$ and $a_{\text{min}}$ are the maximum and minimum acceleration rates, respectively. Then, the acceleration rates $\hat{a}_{i,k}, \hat{a}_{f,k}$, and $\hat{a}_{f\prime,k}$ in Eq.~(\ref{eq:Atr_acc}) can be computed by Eq.~(\ref{eq:IDM1}) and (\ref{eq:IDM2}) at time step $k$. 

Then, we define the control objective $J^{\text {acc}}_k$ as:  
\begin{align}
J^{\text {acc}}_k = \varsigma_{k} \max\{0, B_{k}^{\text{acc}}\}\label{eq:cost_acc}
\end{align}
which is only active if lane-changing is chosen, i.e., $\varsigma_{k}=1$. 

\vspace{0.3em}
\noindent \emph{(6) MPC Formulation} \label{sec:mpc_summury}
\vspace{0.3em}  

We can summarize the aforementioned discussion and formulate the embedded optimization problem in MPC at time step $k$ as follows: 
\begin{align}
\mathop{\min}_{\bm{U}_k,\bm{X}_k,\bm{Z}_k,\varsigma_{k}} & \text{Eq.(\ref{eq:obj})}\notag \\
\text{s.t.} \quad 
& \text{Eqs.(\ref{eq:system_dynamics})--(\ref{eq:u_bound}), Eqs.(\ref{eq:safety1})--Eqs.(\ref{eq:safety2})},  \notag \\
&  \bm{x}(k  \mid k) = \bm{x}(k)  
\label{eq:const_initial} \\
& \varsigma_{k} \in \{0,1\} \label{eq:lc_binary}
\end{align} 
where Eq.(\ref{eq:const_initial}) sets the initial condition, and Eq.(\ref{eq:lc_binary}) defines the lane-changing decision for the planning horizon as a binary variable. Note that the above optimization problem is a quadratic optimization problem for each $\varsigma_{k}$ (0 or 1), which can be easily solved with the barrier method. 

\subsection{HV Trajectory Prediction} \label{sec:HV_predictor} \vspace{0.3em}

Note that the MPC-based controller presented in Section~\ref{sec:MPC_formulation} (i.e., Eq.(\ref{eq:const_rep})) requires the trajectories of surrounding vehicles as input. To provide such input, we present an HV trajectory prediction algorithm to predict the trajectories of the surrounding HVs over the planning horizon. Although it has been demonstrated that Graph Neural Networks (GNNs) can yield supreme performance in comparison with other algorithms~\cite{jia2023hdgt,alinezhad2023pishgu}, 
existing methods typically assume fixed graph topologies (i.e., the spatial relationship between vehicles). However, in weaving sections, intensive lane changes make the spatial relationship between vehicles highly dynamic, which can significantly impact the performance of existing methods. To address this issue, we incorporate the information contained in the evolution of graphs into existing GNN-based trajectory prediction algorithms via EvolveGCN, a neural network architecture to handle dynamic graphs. Specifically, we combine EvolveGCN with a universal trajectory prediction network called Pishgu~\cite{alinezhad2023pishgu}, thanks to its satisfactory performance and simplicity of implementation. 

At each time step $k$, SV (CAV with index $i$) collects information from surrounding HVs, denoted by $\mathcal{V}\subseteq \mathcal{M}$. Here, we notice that $\mathcal{M}$ contains both CAVs and HVs, while $\mathcal{V}$ consists of only HVs. Let a complete graph $\mathcal{G} = (\mathcal{V}, \mathcal{E})$ denote the graph representing the topology of the surrounding HVs. Note that for brevity, we omit the subscript $i$ of SV in this subsection, if such omission does not cause ambiguity. 

In graph $\mathcal{G}$, the features of each HV $j \in \mathcal{V}$, denoted by $\bm{\eta}_{j,k}$, contain five channels and span the previous $T_{\text{in}}$ time steps, including HV $j$'s absolute position $\bm{P}_{j,k} = \{\bm{p}_j^{k-h}\}_{h=0}^{T_{\text{in}}-1}$, relative position to SV $\Delta\bm{P}_{j,k} = \{\Delta\bm{p}_j^{k-h}\}_{h=0}^{T_{\text{in}}-1}$, driving speed $\bm{V}_{j,k} = \{v_j^{k-h}\}_{h=0}^{T_{\text{in}}-1}$, heading angle $\bm{\Psi}_{j,k} = \{\psi_j^{k-h}\}_{h=0}^{T_{\text{in}}-1}$, and lane change intention $\bm{\varsigma}_{j,k} = \{\varsigma_j^{k-h}\}_{h=0}^{T_{\text{in}}-1}$. Notice that the lane change intention of surrounding HVs can be inferred from the state of the turn signal. 
The feature set of all HV neighbors of SV, $\bm{\eta}_{k}$, can be represented by $ \bm{\eta}_{k} = \{\bm{\eta}_{j,k}\}_{j \in \mathcal{V}}$.
The prediction algorithm predicts the future trajectory of all the nodes in graph $\mathcal{G}$, which is denoted as $\hat{\bm{P}}_{k} = \{\hat{\bm{P}}_{j,k}\}_{j \in \mathcal{V}}$, where ${\hat{\bm{P}}_{j,k}} = \{\hat{\bm{p}}_j^{k+h}\}_{h=1}^{P}$ within the prediction horizon of MPC. 

\begin{figure}[h]
	\centering
	\includegraphics[width=0.48\textwidth]{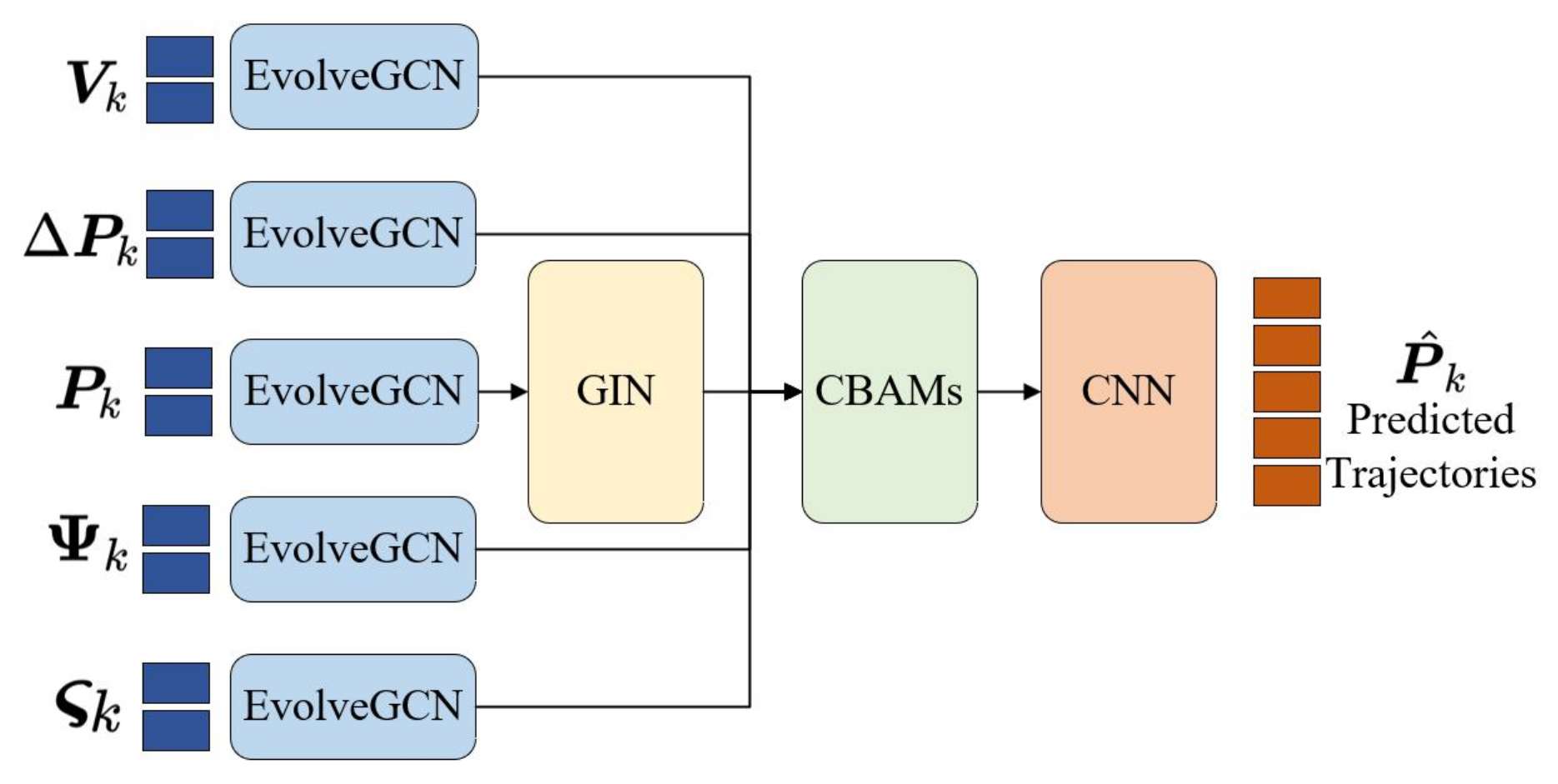}
	\caption{Framework of the trajectory prediction algorithm.}
	\label{fig:predictor}
\end{figure}

Fig.~\ref{fig:predictor} shows the framework of the trajectory prediction algorithm, which comprises four modules with different architectures: EvolveGCN, Graph Isomorphism Network (GIN), Convolutional Block Attention Modules (CBAM), and CNN. 
Note that in addition to the original Pishgu model solely relying on GIN, CBAM, and CNN, we integrate EvolveGCN to account for graph evolution. 

\begin{figure}[h]
	\centering
	\includegraphics[width=0.48\textwidth]{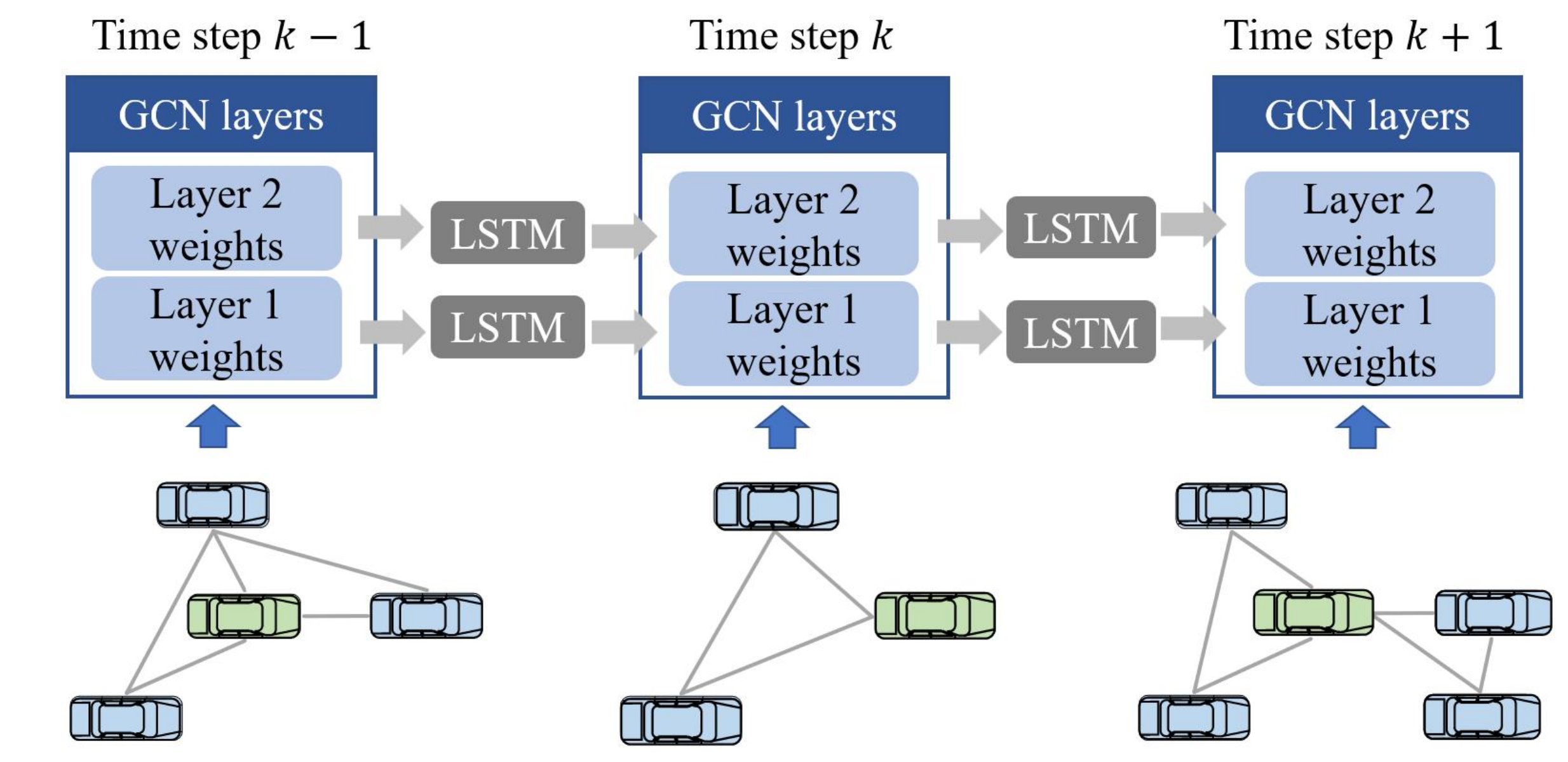}
	\caption{Principle diagram of EvolveGCN layer. The green vehicle represents SV and blue vehicles indicates SV's surrounding HVs within the communication range.}
	\label{fig:detailed_EG}
\end{figure}

\vspace{0.3em}
\noindent \emph{(i) EvolveGCN}. We design an EvolveGCN-based input layer to preprocess the features of all HV consisting of five channels, i.e., $\bm{V}_{k}, \Delta\bm{P}_{k}, \bm{P}_{k}, \bm{\Psi}_{k}, \bm{\varsigma}_{k}$, to capture the evolution of graph $\mathcal{G}$. 
As depicted in Fig.~\ref{fig:detailed_EG}, EvolveGCN dynamically captures the evolution of graph topology by employing a Long Short-Term Memory (LSTM) to iteratively adjust the parameters of the Graph Convolutional Network (GCN). In EvolveGCN, the weight matrix of GCN recursively serves as both the input and the output of the LSTM over time. 
This allows for the direct adaptation of the GCN model across the temporal dimension without relying on the assumption of a fixed number of nodes, while also considering the evolution of the graph topology through the updating of the GCN weight matrix.

\vspace{0.3em}
\noindent \emph{(ii) GIN}. After feeding the features to EvolveGCN, we use GIN to further process the feature of absolute positions $\bm{P}_{k}$ to capture the strong temporal correlation between past absolute positions and future absolute positions. 
The benefit of using GIN instead of GCN is that by introducing self-loop edges into graph convolution, GIN can emphasize and prioritize the ego node's position over information from surrounding nodes. 
Note that GIN processes only the feature of absolute positions to highlight the temporal correlation between past and future absolute positions. 
This design allows the absolute position to play a dominant role in the feature processing process. 

\vspace{0.3em}
\noindent \emph{(iii) CBAM}.
The CBAM layer comprises a channel-spatial attention layer followed by a succeeding CNN layer. Here, the attention mechanism enables the CNN layer to dynamically determine the importance and contribution of the preceding modules' output features to the final prediction. Specifically, channel attention identifies important channels (e.g., absolute positions, speed, etc.), and spatial attention locates the important nodes of the graph. In the established model, the CBAM module is designed to consist of three sequential CBAM layers. 

\vspace{0.3em}
\noindent \emph{(iv) CNN}.
Finally, the output of the last CBAM layer passes through a conventional CNN layer to generate the final predicted trajectories $\hat{\bm{P}}_{k}$. Here, CNN is chosen for its compatibility and high performance in various prediction tasks with multimodal features~\cite{marchetti2020mantra,liao2024human}.

We train the neural network by minimizing the $l_2$-loss, written as 
\begin{align}
    \mathcal{L}_2(\tilde{\bm{P}}_{k}, \hat{\bm{P}}_{k}) = \|\tilde{\bm{P}}_{k}- \hat{\bm{P}}_{k}\|_2^2
\end{align}
where $\tilde{\bm{P}}_{k}$ represents the real trajectory of SV's surrounding HVs in the training data.

\section{Simulation} \label{sec:Result} 
In this section, we perform simulations to evaluate the proposed bi-level controller in a typical weaving section. Section~\ref{sec:benchmark} introduces the 
benchmark algorithms, Section~\ref{sec:implement} describes the simulated scenario, Section~\ref{sec:testing} evaluates the performance of the proposed algorithm, and Section~\ref{sec:sens_com} performs sensitivity analysis to the cell size and upper-level decision-making frequency. 

\subsection{Benchmarks} \label{sec:benchmark}

The following control methods are compared to evaluate the proposed method:
\begin{itemize}
    \item HDM: This benchmark, i.e., human driver model (HDM), implements a rule-based controller without coordination, which uses IDM for longitudinal control and MOBIL for lateral control. This benchmark resembles the driving behavior of HVs.
    \item HLC-L: This benchmark represents a rule-based controller with local coordination. Specifically, this benchmark replicates the work of~\cite{wang2021harmonious}, presenting a harmonious lane-changing (HLC) controller designed to coordinate vehicles within a limited range. 
    \item HLC-G: This benchmark represents a rule-based controller with global coordination. Specifically, this benchmark is an extension of the work~\cite{wang2021harmonious} with local coordination extended to the global scale, which is achieved by employing an EvolveGCN-based RL algorithm to centrally coordinate all vehicles in the weaving section. 
    \item MPC only: This benchmark presents an MPC-based controller with no coordination, which is implemented solely based on our proposed lower-level MPC controller to improve ego travel efficiency. 
    \item Bi-Level-L: This benchmark presents an MPC-based controller with local coordination, which implements our proposed bi-level controller with only local information as in HLC-L controller. 
    \item Bi-Level-G: This benchmark implements the complete version of our proposed bi-level controller, i.e., an MPC-based controller with global coordination.
\end{itemize}
The comparison between MPC, Bi-Level-L, and Bi-Level-G (as well as HDM, HLC-L, and HLC-G) illustrates the value of performing global coordination, while the comparison between MPC-based methods and HLC-G demonstrates the value of using lower-level MPC. 

\subsection{Implementation Details} \label{sec:implement}

We conduct a simulation-based case study~\cite{highway-env}, inspired by a weaving section near Basel, Switzerland, where the parameters are calibrated with empirical data on Nov. 3, 2014. The weaving section is 535 meters long and comprises two main lanes and one auxiliary lane. The simulation replicates a typical evening peak from 16:00 to 17:00, where a capacity drop is often observed. We set a time step of 0.2\,s. 

In the upper-level RL controller, we divide each lane of the weaving section evenly into 20 cells. Hence, the cell representation $\mathcal{N}$ of the upper-level RL controller is configured as a $3 \times 60 \times 5$ tensor. These three dimensions correspond to the feature dimension, the number of cells, and the length of historical information, respectively. 
In our reward function, $q_{\text{max}}$ is set to $1800$\,veh/(lane $\cdot$ h) following the setting in~\cite{xue2023platoon}. We also set $\beta_{\text{spd}} = 1, \beta_{\text{flow}} = 1, \beta_{\text{miss}} = -2$ according to~\cite{wang2021harmonious}.
Moreover, the actor network of PPO includes two CNN layers, succeeded by three fully connected layers with 128, 512, and 384 neurons, respectively. An output layer with 300 neurons is added to produce actions for 60 cells. 
Similarly, the critic network of PPO includes two CNN layers, succeeded by three fully connected layers with 64, 128, and 64 neurons, respectively. The output dimension is set to 1 to produce an estimated scalar state value.
The learning rates of actor and critic nets are both set to be 0.0003, which is an empirical value widely used in existing literature~\cite{wei2023bayesian,lin2021softgym}. 

In the lower-level MPC controller, the conventional values of control weights, $\alpha^{\text{effort}}, \alpha^{\text{exit},\text{x}}, \alpha^{\text{exit},\text{y}}, \alpha^{\text{vel}},$ and $\alpha^{\text{acc}}$, are set to be 0.05, 0.25, 0.5, 0.5, and 0.05, respectively, following the suggestions of empirical studies~\cite{dixit2019trajectory,zhang2019game}. Notice that for bi-level controllers (i.e., Bi-Level-G and Bi-Level-L), these coefficients will be controlled by the upper-level controller. Moreover, the prediction horizon of the MPC controller is $P=16$.

In the HV trajectory predictor, the length of the past observation is $T_{\text{in}} = 10$, following the original setting of Pishgu. The learning rate of the trajectory predictor is 0.0005. 
We generate the HV trajectory training dataset by randomly generating 100 simulation scenarios with different penetration rates of CAV of 100\%, 80\%, 60\%, 40\%, and 20\%. Within each scenario, we collect the real-time locations, speeds, heading angles, and lane change control variables of each vehicle. In total, we collect 530,000 samples, which are then transformed into the training data of the neural networks.  

Each trainable benchmark (Bi-Level-G, Bi-Level-L, HLC-G, and HLC-L) is trained for 50 episodes, during which they converge to optimal performance. Then an additional five episodes are conducted to test the control performance. The experiments are conducted on a computer with an Intel i7-12700 CPU@2.10 GHz. The designed neural networks and MPC control scheme are implemented in Python with Pytorch 1.13.1 and Pyomo 6.6.1. 

\subsection{Performance in testing scenarios} \label{sec:testing}

\begin{figure*}[hbp]
    \centering
    
    \begin{subfigure}[b]{0.47\textwidth}
        \includegraphics[width=0.8\textwidth]{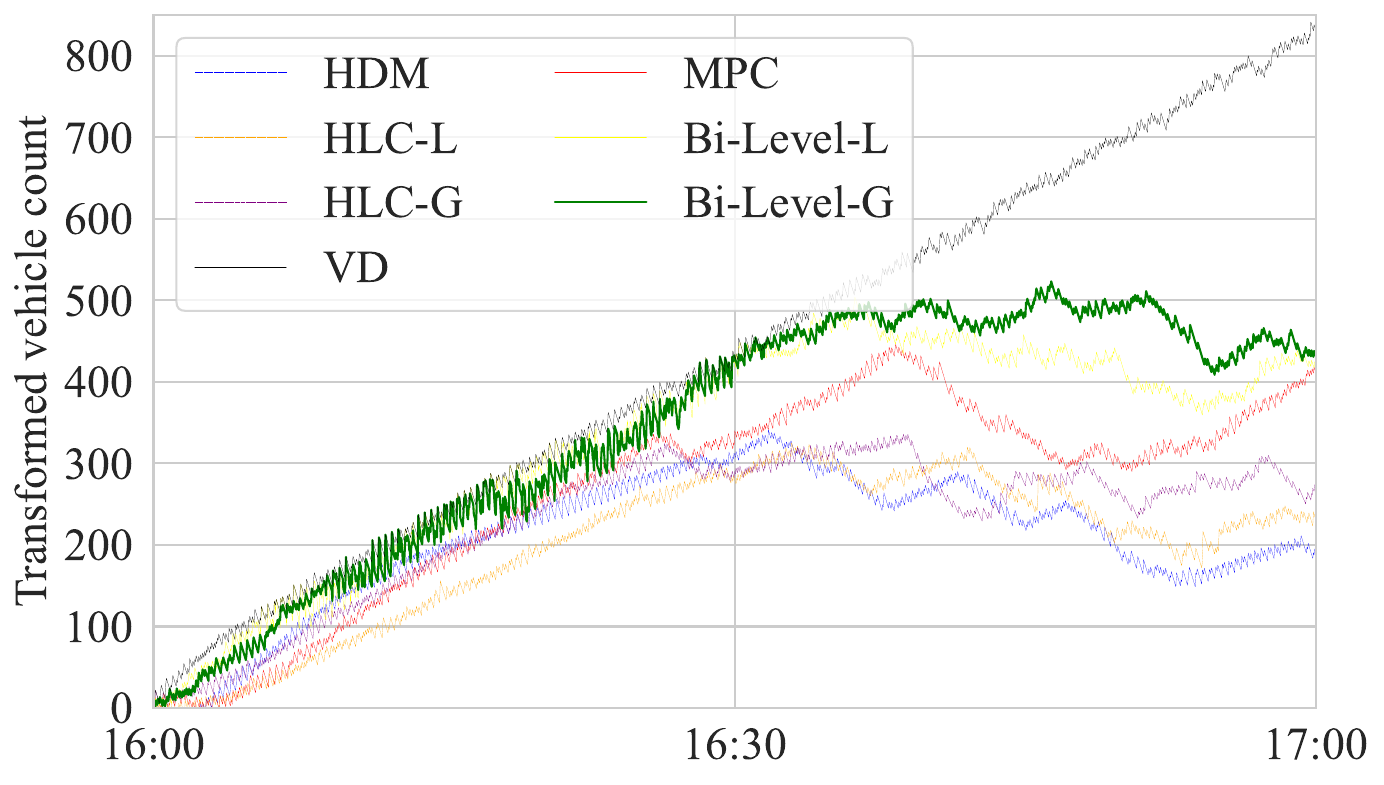}
        \caption{Queuing diagram under 20\% CAV penetration rate.}
    \end{subfigure} 
    \begin{subfigure}[b]{0.44\textwidth}
        \includegraphics[width=0.8\textwidth]{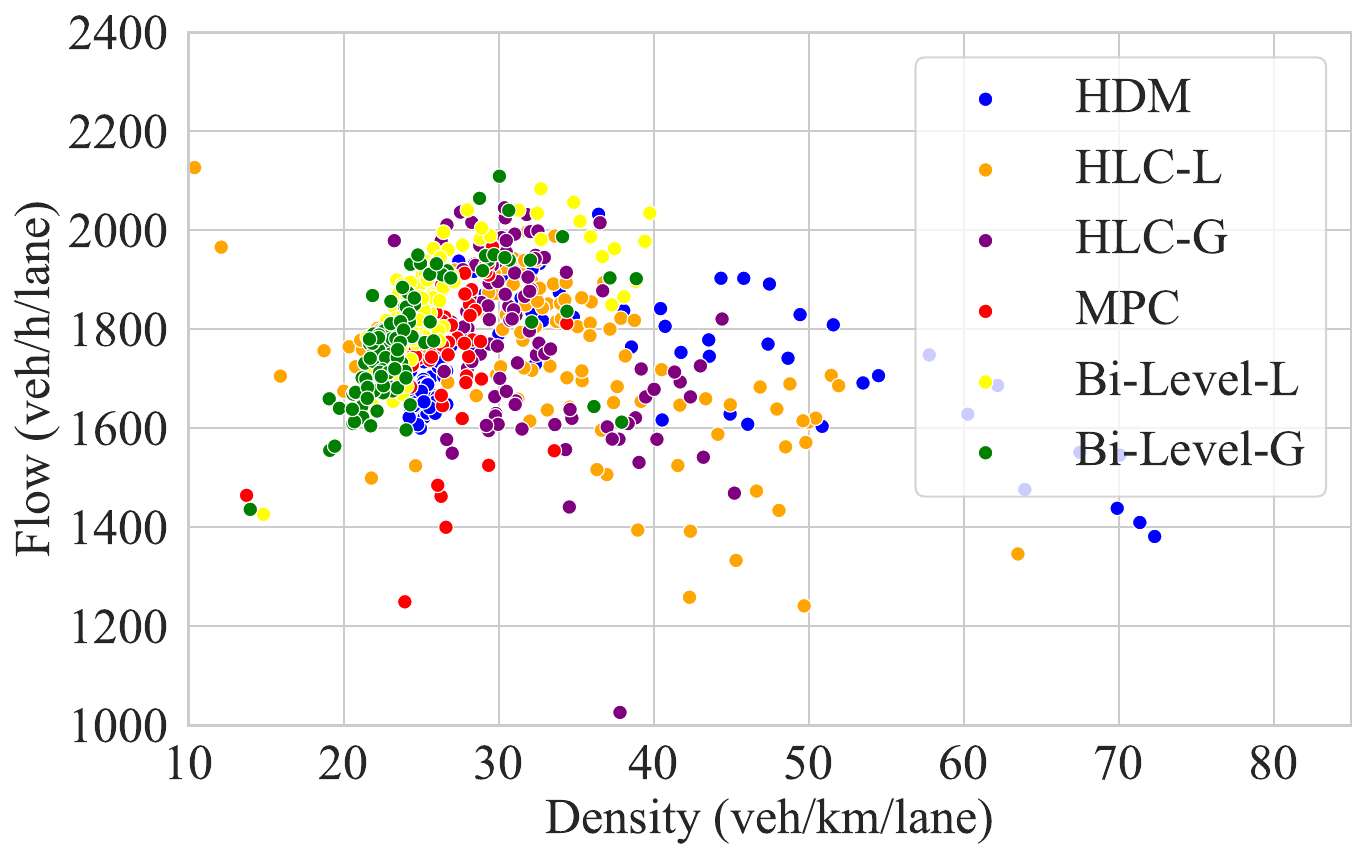}
        \caption{Fundamental diagram under 20\% CAV penetration rate.}
    \end{subfigure} 

        \begin{subfigure}[b]{0.47\textwidth}
        \includegraphics[width=0.8\textwidth]{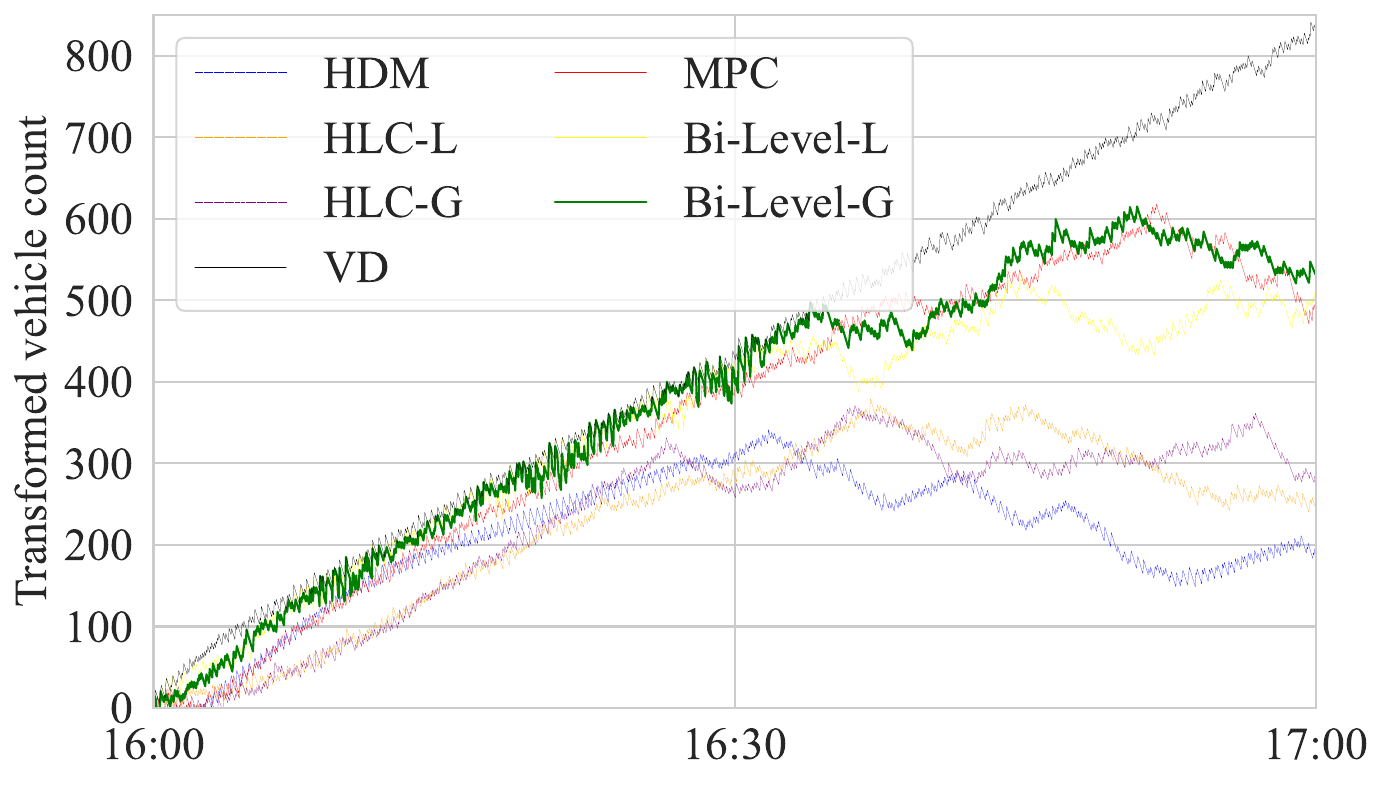}
        \caption{Queuing diagram under 40\% CAV penetration rate.}
    \end{subfigure} 
    \begin{subfigure}[b]{0.44\textwidth}
        \includegraphics[width=0.8\textwidth]{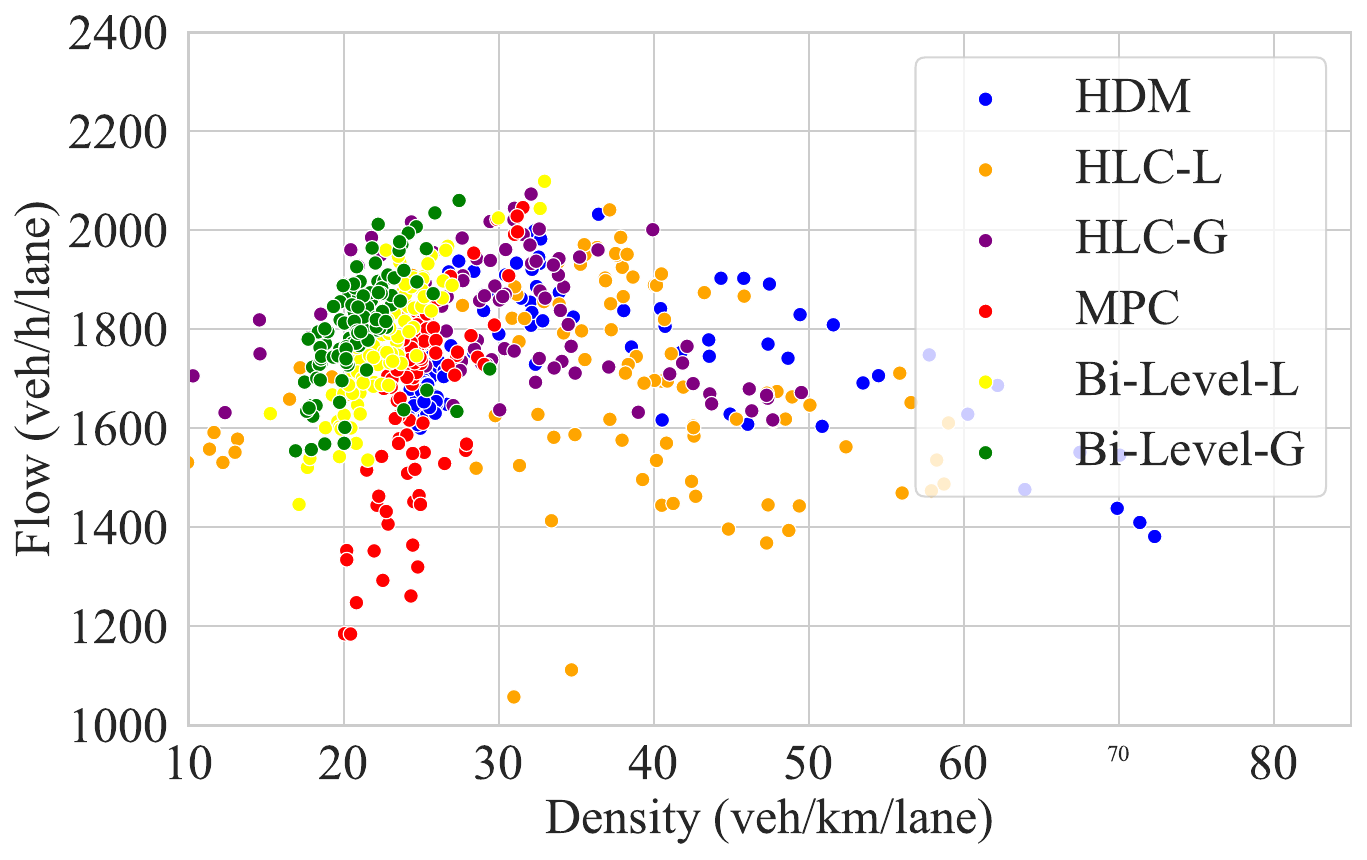}
        \caption{Fundamental diagram under 40\% CAV penetration rate.}
    \end{subfigure} 

    \begin{subfigure}[b]{0.47\textwidth}
        \includegraphics[width=0.8\textwidth]{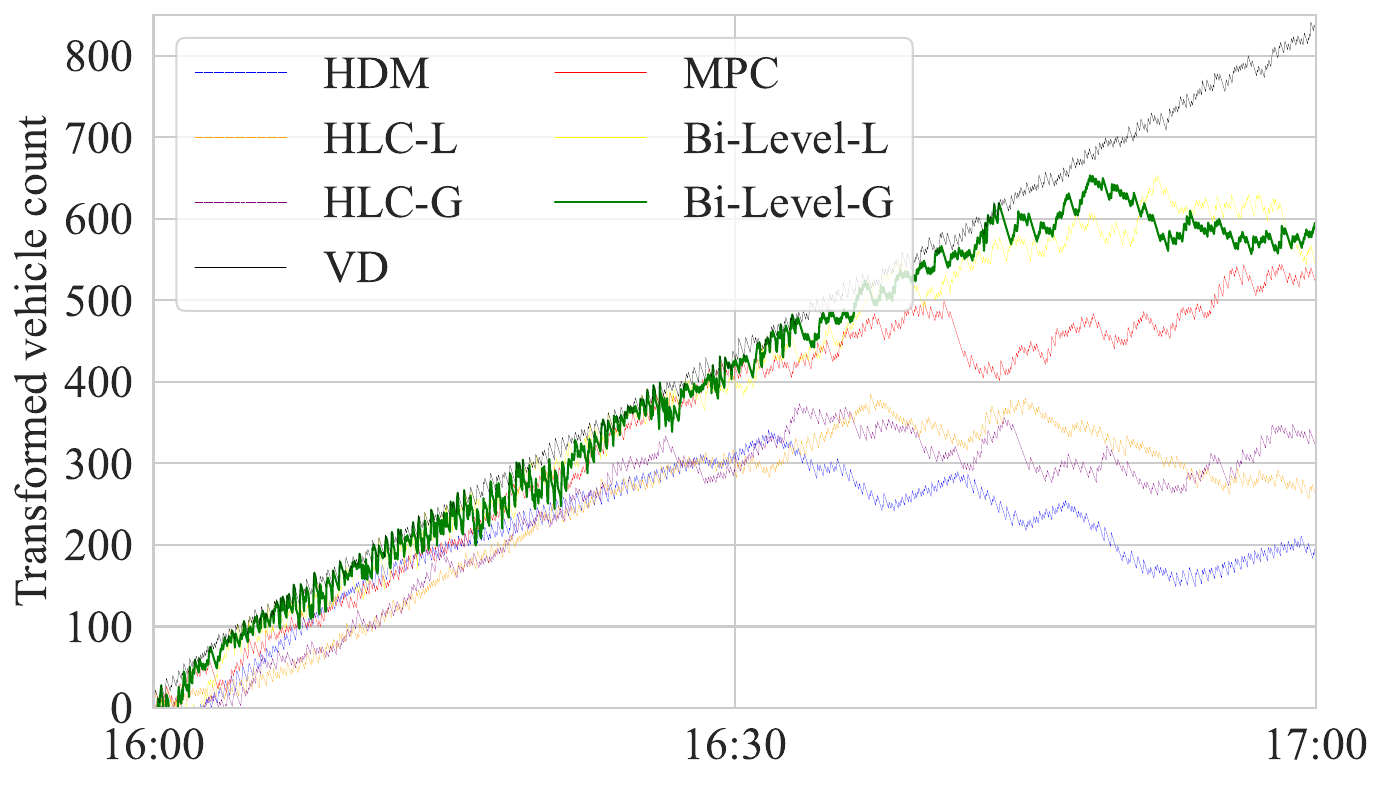}
        \caption{Queuing diagram under 60\% CAV penetration rate.}
    \end{subfigure} 
    \begin{subfigure}[b]{0.44\textwidth}
        \includegraphics[width=0.8\textwidth]{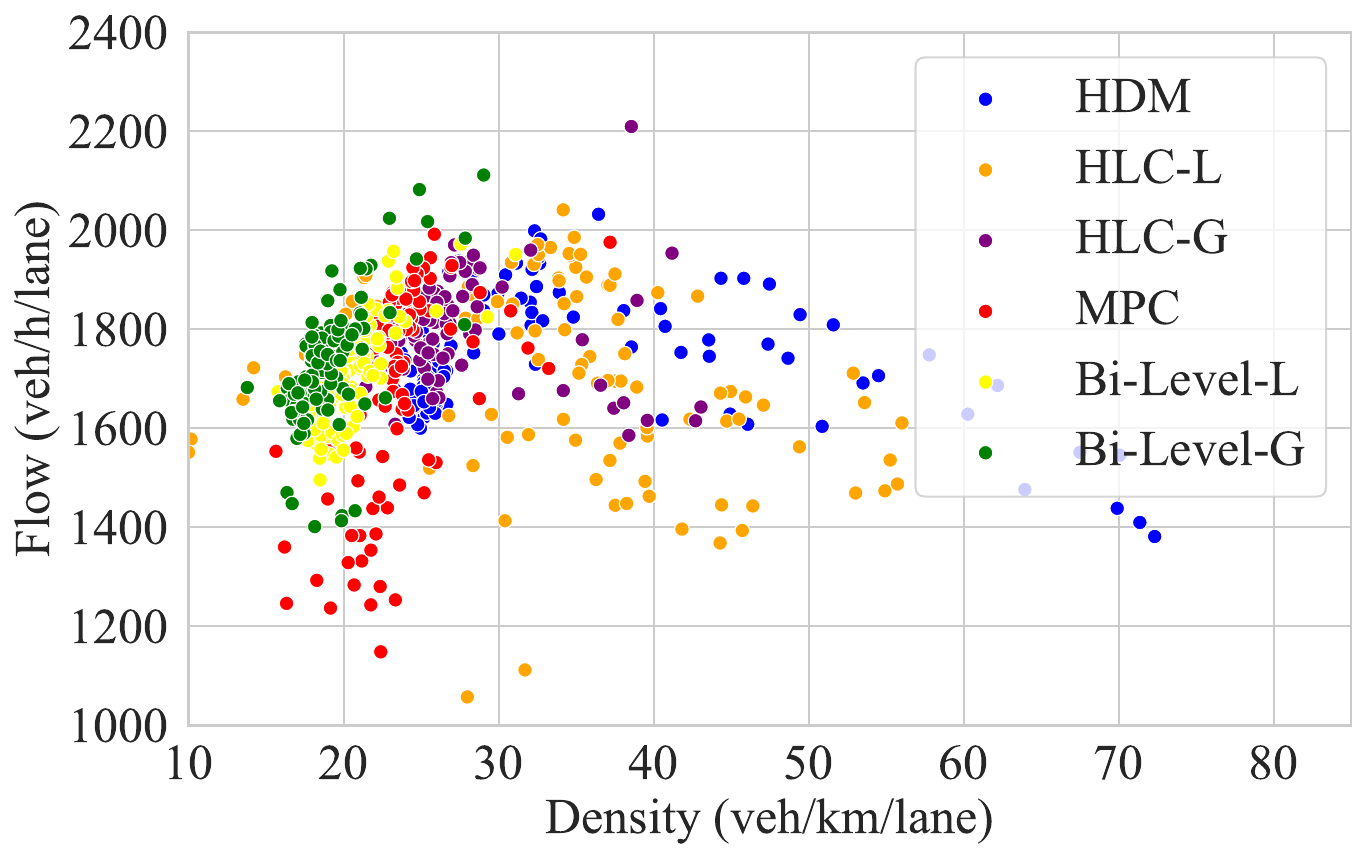}
        \caption{Fundamental diagram under 60\% CAV penetration rate.}
    \end{subfigure} 

    \begin{subfigure}[b]{0.47\textwidth}
        \includegraphics[width=0.8\textwidth]{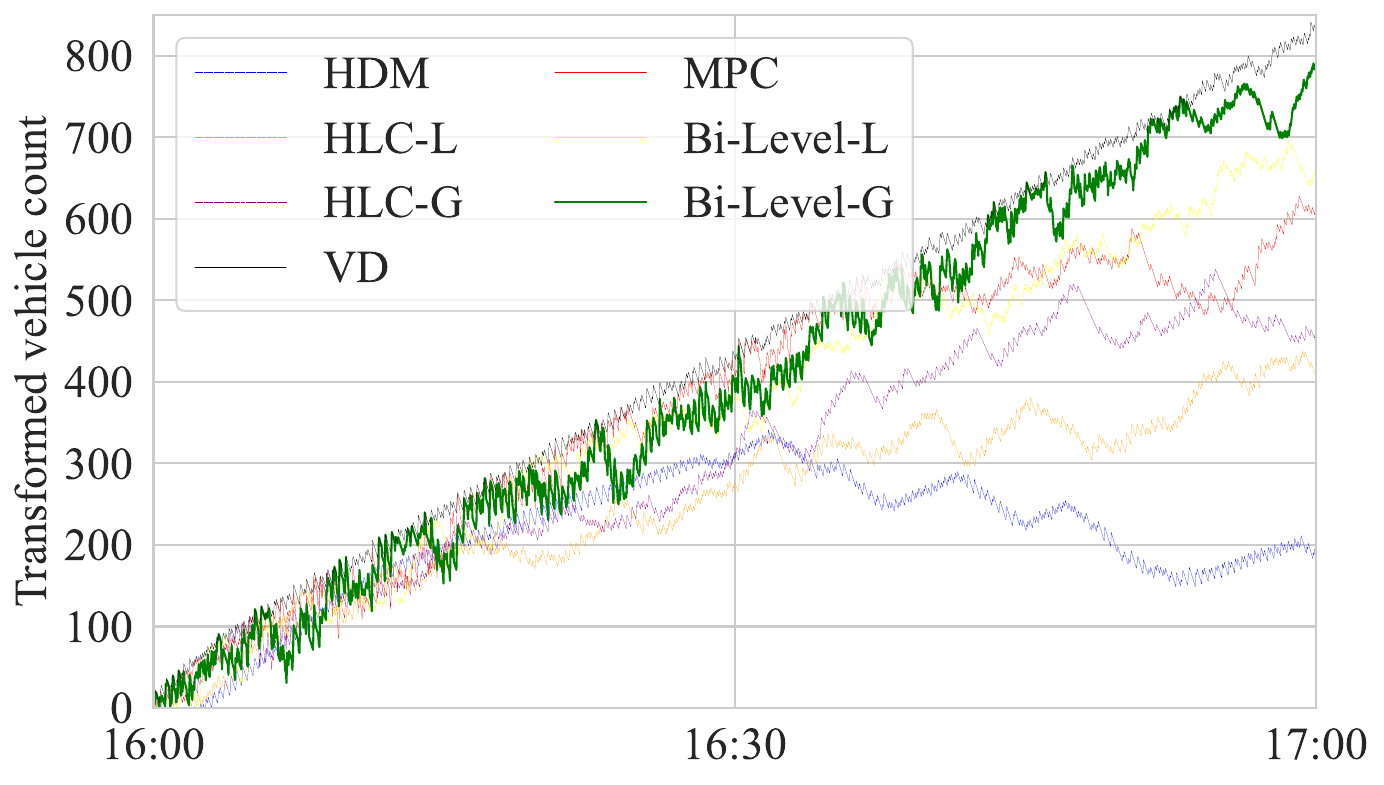}
        \caption{Queuing diagram under 80\% CAV penetration rate.}
    \end{subfigure}
    \begin{subfigure}[b]{0.44\textwidth}
        \includegraphics[width=0.8\textwidth]{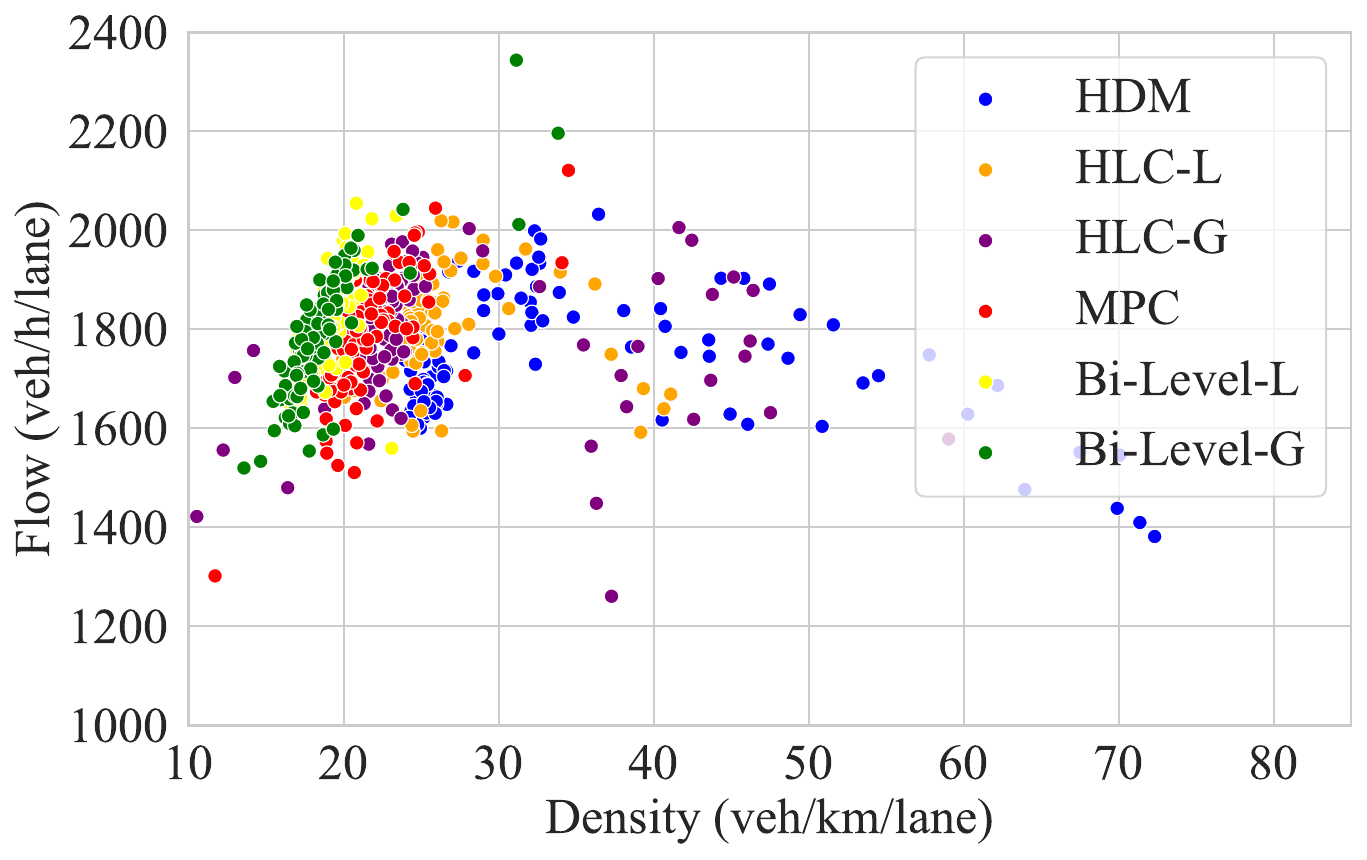}
        \caption{Fundamental diagram under 80\% CAV penetration rate.}
    \end{subfigure}  

    \begin{subfigure}[b]{0.47\textwidth}
        \includegraphics[width=0.8\textwidth]{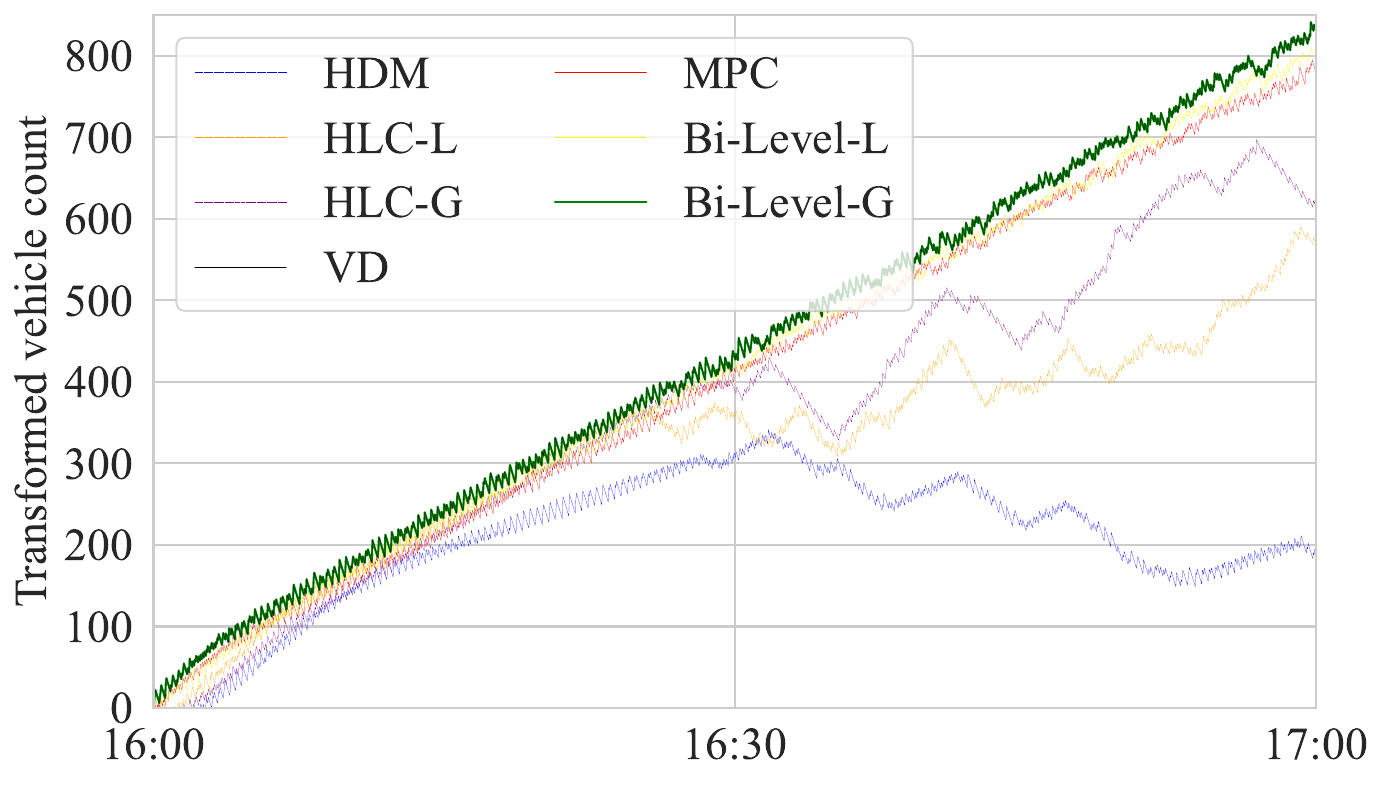}
        \caption{Queuing diagram under 100\% CAV penetration rate.}
    \end{subfigure}
    \begin{subfigure}[b]{0.44\textwidth}
        \includegraphics[width=0.8\textwidth]{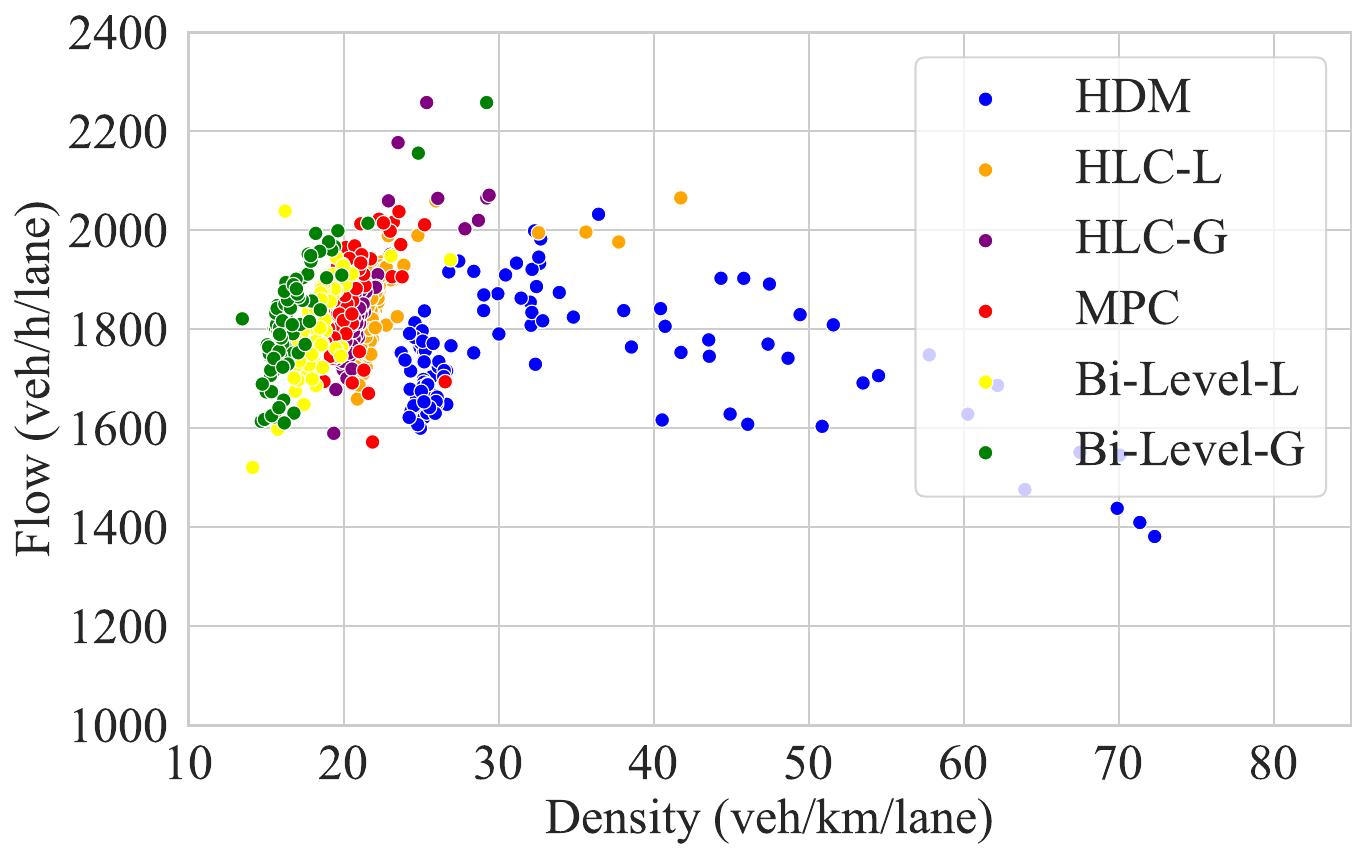}
        \caption{Fundamental diagram under 100\% CAV penetration rate.}
    \end{subfigure}
    
    \caption{Queuing diagrams with background flow of 4800 veh/h and fundamental diagrams under different CAV penetration rates. Notice that VD represents the virtual departure curve at the exit of the weaving section, which is obtained from the simulated traffic demand based on the empirical data.}
    \label{fig:que_FD_diag}
\end{figure*}

\begin{table}[htp]
    \centering
    \caption{Benchmark performance in scenarios with various CAV penetration rates. The values in parentheses indicate the percentage of improvement compared to the HDM benchmark. The optimal algorithm for each CAV penetration rate is highlighted in bold.}
    \label{tab:numerical_com}
   
    \begin{tabular}{p{0.5in}p{0.7in}p{0.87in}p{0.8in}}
    \toprule
    \textbf{Penetration Rate} & \textbf{Benchmark} & \textbf{Space Mean Speed (km/h)} & \textbf{Exit Flow (veh/(lane $\cdot$ h))} \\ 
    
    \midrule
    \multirow{1}{*}{0\%} 
    & HDM & 55.45 & 1663 \\
    
    \cmidrule(lr){1-4}
    \multirow{4}{*}{20\%} 
    & HLC-L & 56.92 (2.65\%) & 1677 (0.84\%)  \\
    & HLC-G & 57.04 (2.86\%) & 1690 (1.62\%)  \\
    & MPC & 57.97 (4.54\%) & 1737 (4.44\%)  \\
    & Bi-level-L & 58.80 (6.04\%) & 1740 (4.63\%)   \\
    & \textbf{Bi-Level-G} & \textbf{60.35 (8.83\%)} & \textbf{1743 (4.81\%)} \\
    
    \cmidrule(lr){1-4}
    \multirow{4}{*}{40\%} 
    & HLC-L & 58.79 (6.02\%) & 1682 (1.14\%)   \\
    & HLC-G & 59.02 (6.43\%) & 1694 (1.86\%)   \\
    & MPC & 60.20 (8.56\%) & 1762 (5.95\%)     \\
    & Bi-level-L & 62.02 (11.84\%) & 1769 (6.37\%)   \\
    & \textbf{Bi-Level-G} & \textbf{63.67 (14.82\%)} & \textbf{1777 (6.85\%) }  \\
    
    \cmidrule(lr){1-4}
    \multirow{4}{*}{60\%} 
    & HLC-L & 59.79 (7.82\%) & 1687 (1.44\%)  \\
    & HLC-G & 60.30 (8.74\%) & 1707 (2.64\%)  \\
    & MPC & 63.89 (15.22\%) & 1774 (6.67\%)  \\
    & Bi-level-L & 64.72 (16.71\%) & 1780 (7.03\%)   \\
    & \textbf{Bi-Level-G} & \textbf{66.39 (19.72\%)} & \textbf{1797 (8.05\%) }  \\
    
    \cmidrule(lr){1-4}
    \multirow{4}{*}{80\%} 
    & HLC-L & 63.23 (14.03\%) & 1735 (4.32\%)   \\
    & HLC-G & 64.85 (16.95\%) & 1753 (5.41\%)   \\
    & MPC & 70.80 (27.68\%) & 1804 (8.47\%)   \\
    & Bi-level-L & 71.34 (28.65\%) & 1817 (9.26\%)   \\
    & \textbf{Bi-Level-G} & \textbf{74.22 (33.85\%)} & \textbf{1860 (11.84\%) }  \\
    
    \cmidrule(lr){1-4}
    \multirow{4}{*}{100\%} 
    & HLC-L & 74.09 (33.61\%) & 1790 (7.63\%)   \\
    & HLC-G & 76.17 (37.36\%) & 1809 (8.77\%)   \\
    & MPC & 80.11 (44.47\%) & 1860 (11.84\%)    \\
    & Bi-level-L & 83.15 (49.95\%) & 1867 (12.26\%)   \\
    & \textbf{Bi-Level-G} & \textbf{85.80 (54.73\%)} & \textbf{1877 (12.86\%)} \\
    
    \bottomrule
    \end{tabular}
\end{table}

\begin{figure}[h]
	\centering
	\includegraphics[width=0.42\textwidth]{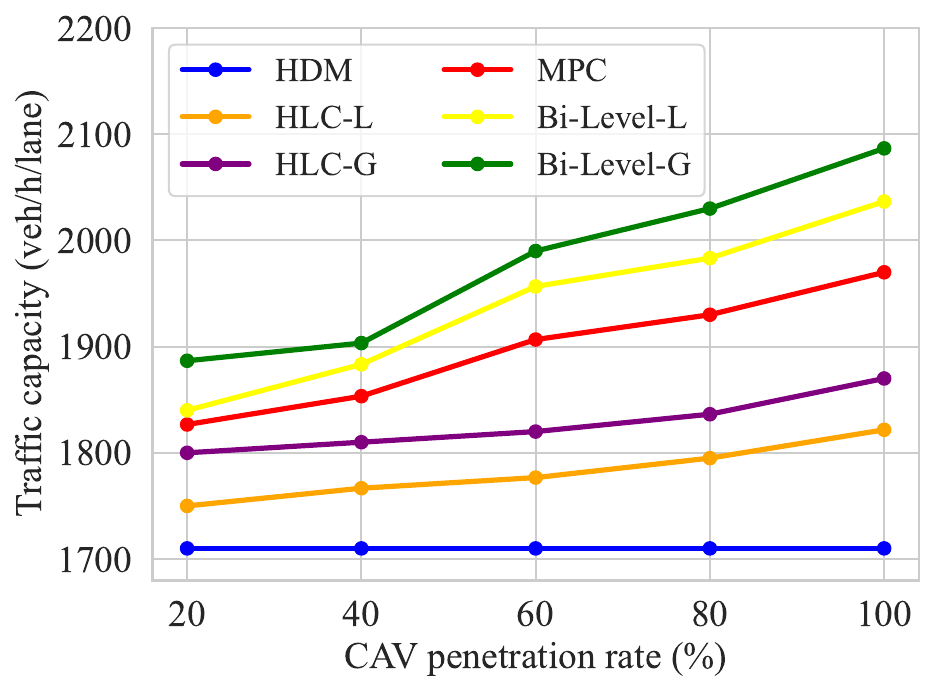}
	\caption{Traffic capacity comparison under different CAV penetration rates.}
	\label{fig:cap_com}
\end{figure}

Table~\ref{tab:numerical_com} and Fig.~\ref{fig:que_FD_diag} show the performance of the benchmarks in scenarios with various CAV penetration rates, where the incoming flow to the weaving section is obtained from empirical data. Specifically, Table~\ref{tab:numerical_com} summarizes the space-mean speed and exit flow of the weaving section resulting from all benchmarks, and Fig.~\ref{fig:que_FD_diag} demonstrates the queuing diagrams and fundamental diagrams resulting from all benchmarks.  
Fig.~\ref{fig:cap_com} evaluates the resulting capacities from all benchmarks in scenarios with various CAV penetration rates, which are obtained by increasing the incoming flow to the weaving section until a sustainable outflow can be measured.

\vspace{0.3em}
\noindent \emph{Value of global coordination}. We show the value of coordination by comparing Bi-Level-G with Bi-Level-L and MPC controllers.
First, from Table~\ref{tab:numerical_com}, we can see a continuous improvement in the space-mean speed and exit flow, as the level of coordination increases. Specifically, Bi-Level-G consistently achieves the highest space-mean speeds and exit flow in scenarios with various CAV penetration rates, outperforming both Bi-Level-L and MPC controllers. Additionally, Bi-Level-L also yields better performance than MPC. 
Second, we can see from the queuing diagrams (i.e., the left column) of Fig.~\ref{fig:que_FD_diag} that Bi-Level-G reduces the magnitude of the capacity drop in all scenarios tested, especially in scenarios with low penetration rates (e.g., 20\% as in Fig.~\ref{fig:que_FD_diag}a). 
Notice that under 100\% CAV penetration rate, the cumulative curve corresponding to Bi-Level-G completely overlaps with the virtual departure curve, showing that our coordination algorithm can significantly mitigate congestion.
We can also see from the fundamental diagrams (i.e., the right column) of Fig.~\ref{fig:que_FD_diag} that Bi-Level-G improves the slope of the free-flow branch, i.e., enabling a higher driving speed. Third, we can see from Fig.~\ref{fig:cap_com} that the capacity increases as the level of coordination increases from no coordination to global coordination, i.e., Bi-Level-G yields a higher capacity than both Bi-Level-L and MPC.  
Similar conclusions can also be drawn from rule-based controllers by comparing HDM, HLC-L, and HLC-G.  

\vspace{0.3em}
\noindent \emph{Value of Lower-level Controller}. 

We assess the value of our lower-level MPC controller by comparing the MPC controller with the HLC-G controller. As shown in Table~\ref{tab:numerical_com}, it is evident that all MPC-based controllers (Bi-Level-G, Bi-Level-L and MPC) outperform the rule-based controllers (HLC-G, HLC-L and HDM) in terms of space-mean speed and exit flow. 
We can also see from Fig.~\ref{fig:que_FD_diag} and  Fig.~\ref{fig:cap_com} that MPC-based approaches significantly reduce the magnitude of capacity drop,  achieve a higher driving speed, and improve capacity, compared to rule-based controllers.  

\vspace{0.3em}
\noindent \emph{Value of CAV Penetration Rate}. \label{sec:val_CAV} 
We evaluate the value of the CAV penetration rates by comparing benchmark performance under different CAV penetration rates. 
As demonstrated in Table~\ref{tab:numerical_com}, the improvement of Bi-Level-G in terms of space-mean speed and exit flow increases as the CAV penetration rate increases. Specifically, the exit flow can be improved by up to 12.86\%, and the space-mean speed can be boosted by up to 54.73\%. 
Moreover, from Fig.~\ref{fig:cap_com}, we can see a notable increase in capacity as CAV penetration improves, which is valid for any benchmark. 

\vspace{0.3em}
\noindent \emph{Value of EvolveGCN in Trajectory Prediction}.
We evaluate the benefits of incorporating the EvolveGCN layer in the trajectory predictor by comparing the resulting Root Mean Square Errors (RMSEs) of (1) the Pishgu algorithm, (2) the Pishgu algorithm with GCN to handle the spatial correlation between vehicles, and (3) our proposed trajectory prediction algorithm. As shown in Table~\ref{tab:pred_acc}, 
our predictor outperforms the original Pishgu and revised Pishgu with GCN by explicitly considering the dynamic graph topology. 

\begin{table}[htbp]
\caption{Trajectory prediction accuracy (meter)}
\centering
\begin{tabular}{p{1.0cm}|p{0.8cm}|p{0.8cm}|p{0.8cm}|p{0.8cm}|p{0.8cm}|p{0.8cm}}
\hline\hline
\multirow{2}{1.2cm}{Time Step} & \multicolumn{3}{c|}{X-axis} & \multicolumn{3}{c}{Y-axis} \\ \cline{2-7}
 & Pishgu & GCN-Pishgu & \textbf{Ours} & Pishgu & GCN-Pishgu & \textbf{Ours}     \\\hline
1  & 0.253     & 0.206 & 0.191     & 0.016         & 0.014 & 0.012         \\\hline
2  & 0.260     & 0.218 & 0.227     & 0.016         & 0.014 & 0.012         \\\hline
3  & 0.288     & 0.219 & 0.214     & 0.016         & 0.014 & 0.012       \\\hline
4  & 0.288     & 0.231 & 0.201     & 0.016         & 0.014 & 0.012        \\\hline
5  & 0.301     & 0.245 & 0.216     & 0.016         & 0.014 & 0.012        \\\hline
6  & 0.316     & 0.277 & 0.250     & 0.016         & 0.014 & 0.012        \\\hline
7  & 0.387     & 0.345 & 0.311     & 0.016         & 0.014 & 0.013        \\\hline
8  & 0.477     & 0.434 & 0.389     & 0.017         & 0.014 & 0.013        \\\hline
9  & 0.586     & 0.543 & 0.493     & 0.017         & 0.015 & 0.013        \\\hline
10 & 0.714     & 0.669 & 0.611     & 0.017         & 0.015 & 0.013        \\\hline
11 & 0.857     & 0.811 & 0.752     & 0.017         & 0.015 & 0.013        \\\hline
12 & 1.017     & 0.968 & 0.907     & 0.017         & 0.015 & 0.013        \\\hline
13 & 1.191     & 1.139 & 1.082     & 0.017         & 0.015 & 0.014        \\\hline
14 & 1.373     & 1.322 & 1.256     & 0.017         & 0.016 & 0.014        \\\hline
15 & 1.570     & 1.516 & 1.453     & 0.018         & 0.016 & 0.014        \\\hline
16 & 1.772     & 1.718 & 1.661     & 0.019         & 0.017 & 0.015     \\\hline\hline  
\end{tabular}\label{tab:pred_acc}
\end{table}

\subsection{Sensitivity Analysis} \label{sec:sens_cell} 
In this subsection, we perform sensitivity analysis on the cell size and the upper-level decision-making frequency. We specifically compare the resulting traffic capacity from various levels of cell sizes and upper-level decision-making frequencies in scenarios with three typical CAV penetration rates (20\%, 60\%, and 100\%). \vspace{0.3em}

\begin{figure}[h]
	\centering
	\includegraphics[width=0.42\textwidth]{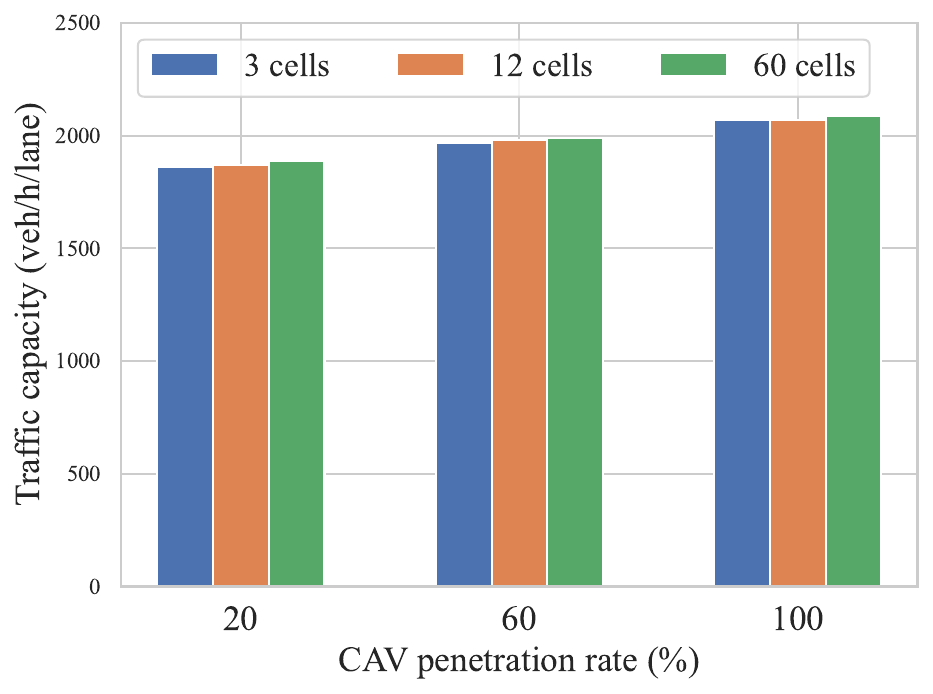}
	\caption{Traffic capacity comparison of Bi-Level-G under different cell sizes.}
	\label{fig:cell_cap}
\end{figure}

\noindent\emph{Sensitivity analysis on cell size}. 
As depicted in Fig.~\ref{fig:cell_cap}, we observe that increasing the cell size slightly improves traffic capacity for all CAV penetration rates. This is because a more refined cell representation enables greater flexibility in tuning control weights. Nevertheless, the improvement is not significant, indicating that we can achieve the benefits of global coordination with a coarse cell representation. 

\begin{figure}[h]
	\centering
	\includegraphics[width=0.42\textwidth]{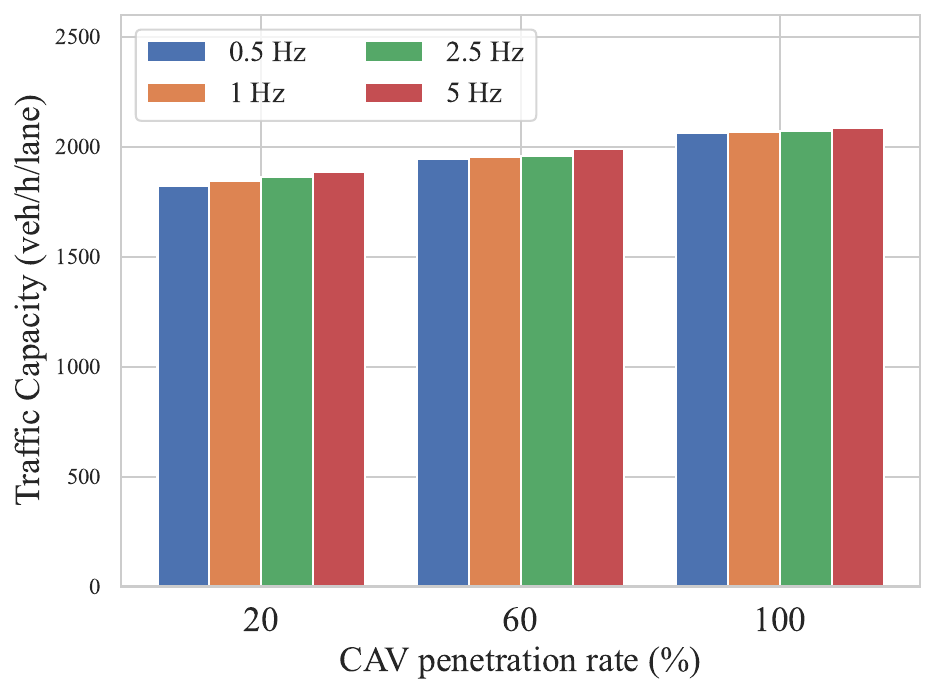}
	\caption{Traffic capacity comparison of Bi-Level-G under different upper-level decision-making frequencies.}
	\label{fig:communicate_cap}
\end{figure}

\vspace{0.3em} \noindent\emph{Sensitivity analysis on upper-level decision-making frequency}. \label{sec:sens_com}
Fig.~\ref{fig:communicate_cap} illustrates the traffic capacity under various decision-making frequencies at the upper level: 0.5Hz, 1Hz, 2.5Hz, and 5Hz. 
The result shows that as the decision-making frequencies at the upper level increases, the traffic capacity improves accordingly in all CAV penetration rates. By increasing the upper-level decision-making frequencies from 0.5Hz to 5Hz, the capacity can increase by up to 3.45\%, which indicates that our proposed framework can be used even if the traffic operators and CAVs are making decisions at different frequencies. 

\section{Conclusion} \label{sec:Conclusion}
In this paper, we propose a bi-level controller to implement global coordinated lane change control in mixed-traffic weaving sections. 
First, the upper level perceives vehicular information from all vehicles and then determines control weights to balance multiple objectives in the lower-level controller. The lower level takes the upper level's decision and solves an optimization problem for each individual vehicle to obtain a decision corresponding to the vehicle's local situation. 
Second, a trajectory prediction algorithm is proposed at the lower level, where EvolveGCN is incorporated to consider the evolution of multiple HVs around a CAV.
The results demonstrate that the proposed algorithm surpasses state-of-the-art baselines, increasing driving speed and traffic flow by up to 54.73\% and 12.86\%, respectively.

This research opens several interesting directions. First, we would like to adapt our proposed approach to other lane-changing scenarios, such as merges and diverges. Second, we would like to improve the generalizability of the proposed approach so that the trained model can be generalized to various weaving sections and/or weather conditions.

\bibliographystyle{IEEEtran}
\input{output.bbl}

\begin{IEEEbiography}
[{\includegraphics[width=1in,height=1.25in,clip,keepaspectratio]{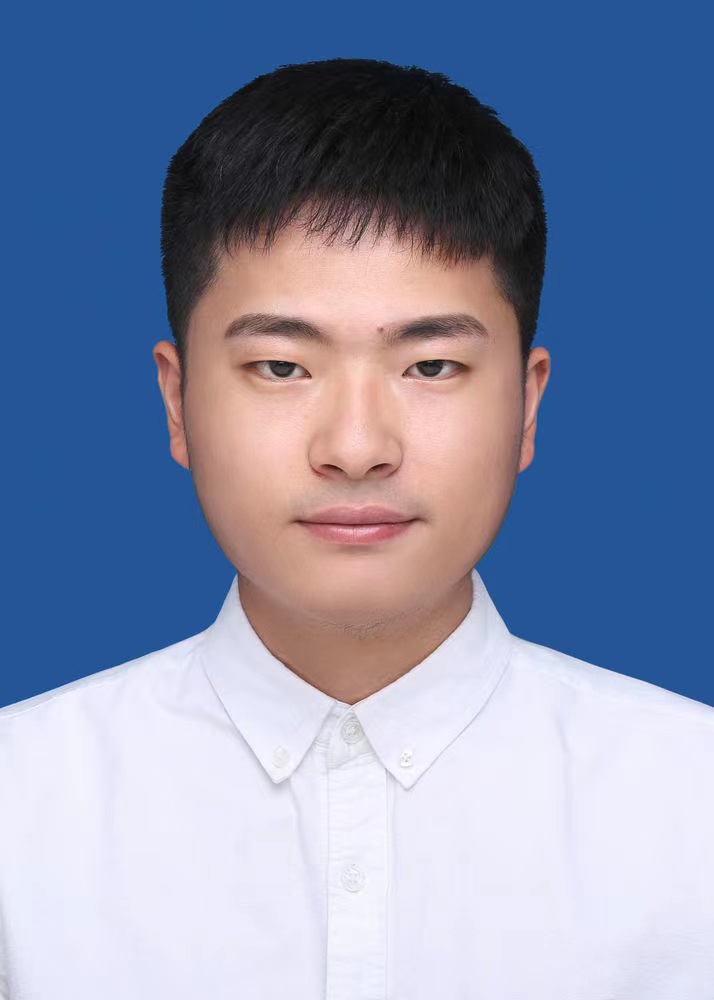}}]{Longhao Yan} receives the B.Eng. degree and M.Eng. degree in School of Electronics and Control Engineering from Chang’an University, Xi’an, China, in 2019 and 2022 respectively. He is currently working towards a Ph.D. degree with the National University of Singapore. His research interests include lateral control and trajectory prediction of intelligent transportation system.
\end{IEEEbiography}
\begin{IEEEbiography}
[{\includegraphics[width=1in,height=1.25in,clip,keepaspectratio]{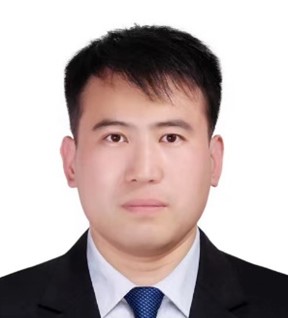}}]{Jinghao Liang} received the B.S. degree from School of Mechanical Engineering, Nanjing University of Science and Technology, Nanjing, China, in 2017, and Ph.D. degree from School of Mechanical Engineering, Southeast University, Nanjing, China, in 2022. Now he is a Research Fellow with Department of Civil and Environmental Engineering, National University of Singapore. His research interests include vehicle dynamics and control, and autonomous vehicles.
\end{IEEEbiography}
\begin{IEEEbiography}
[{\includegraphics[width=1in,height=1.25in,clip,keepaspectratio]{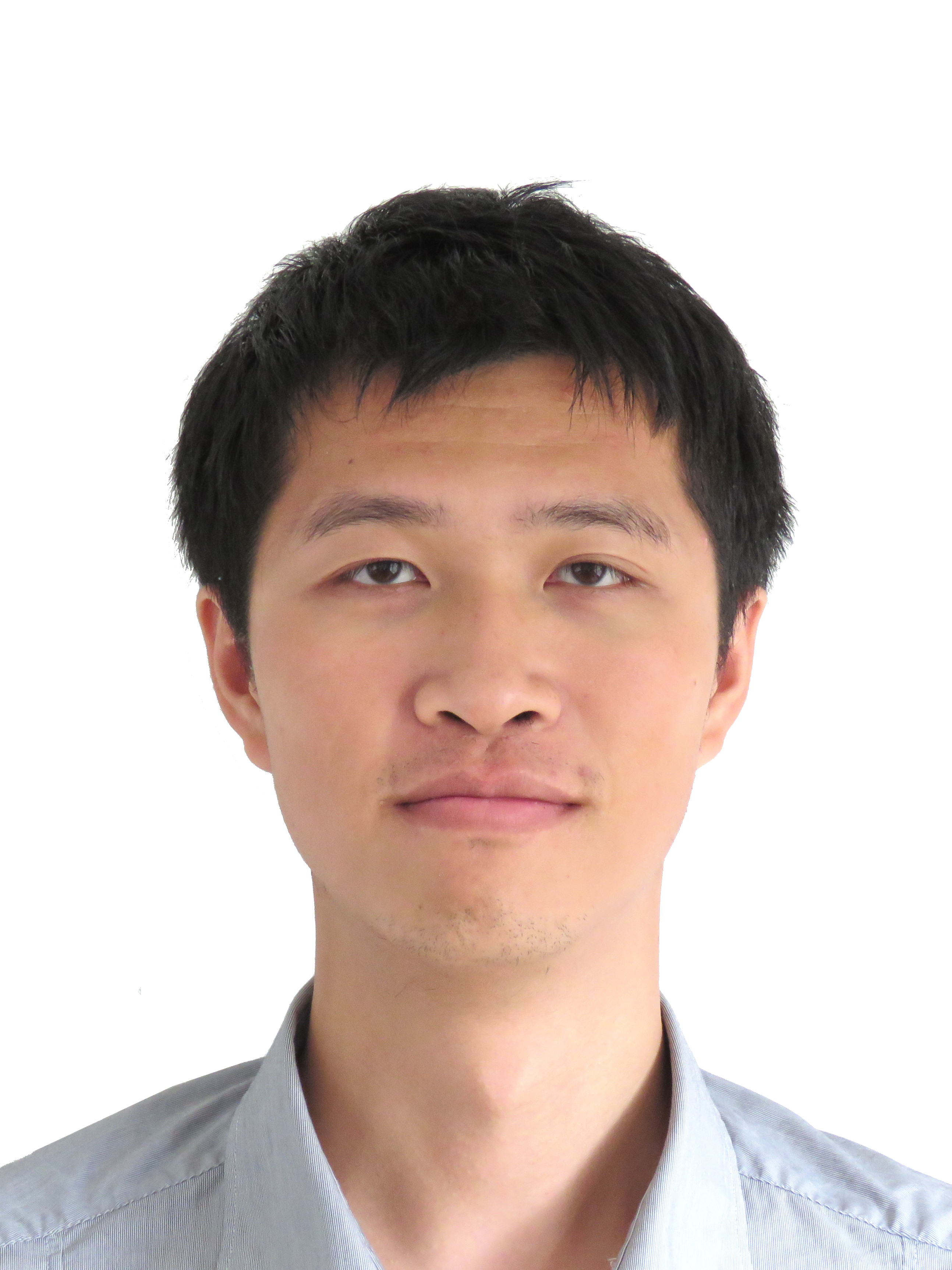}}]{Kaidi Yang} (Member, IEEE) is an Assistant Professor in the Department of Civil and Environmental Engineering at the National University of Singapore. Prior to this, he was a postdoctoral researcher with the Autonomous Systems Lab at Stanford University. He obtained a PhD degree from ETH Zurich and M.Sc. and B.Eng. degrees from Tsinghua University. His main research interest is the operation of future mobility systems enabled by connected and automated vehicles (CAVs) and shared mobility.
\end{IEEEbiography}

\end{document}